# Simultaneous Superconducting and Topological Properties in Mg−Li Electrides at High Pressures

Dan Wang,[1,2] Hongxing Song,[1] Qidong Hao,[1,2] Guangfa Yang, [1] Hao Wang,[1] Leilei Zhang,[3] Ying Chen,[4] Xiangrong Chen,[2*] Huayun Geng[1,5*]

[1] *National Key Laboratory of Shock Wave and Detonation Physics, Institute of Fluid Physics, China Academy of Engineering Physics, Mianyang, Sichuan 621900, P. R. China;*

[2] *Institute of Atomic and Molecular Physics, College of Physics, Sichuan University, Chengdu 610065, P. R. China;*

[3] *Institute of Nano-Structured Functional Materials, Huanghe Science and Technology College, Zhengzhou 450063, P. R. China;*

[4] *Fracture and Reliability Research Institute, School of Engineering, Tohoku University, Sendai 980-8579, Japan;*

[5] *HEDPS, Center for Applied Physics and Technology, and College of Engineering, Peking University, Beijing 100871, P. R. China.*

**Abstract:**

Electrides as a unique class of emerging materials exhibit fascinating properties and hold important significance for understanding the matter under extreme conditions, which is characterized by valence electrons localized into the interstitial space as quasi-atoms (ISQs). In this work, using crystal structure prediction and first-principles calculations, we identified seven stable phases of Mg-Li that are electride with novel electronic properties under high pressure. Among them, $MgLi_{10}$ is a semiconductor with a band gap of 0.22 eV; and *Pm*-3*m* MgLi is superconductor with a superconducting transition temperature of 22.8 K. The important role played by the localization degree of ISQ in the superconducting transition temperature of these electrides is revealed by systematic comparison of Mg-Li with other Li-rich electride superconductors. Furthermore, we proved that *Pm*-3*m* MgLi and *Pnma* MgLi also have distinct topological behavior with metallic surface states and the non-zero $Z_2$ invariant. The simultaneous coexistence of superconductivity, electronic band topology and electride property in the same structure of *Pm*-3*m* MgLi and *Pnma* MgLi demonstrates the feasibility of realizing multi-quantum phases in a single material, which will stimulate further research in these interdisciplinary fields.

* To whom correspondence should be addressed. E-mail: s102genghy@caep.cn; xrchen@scu.edu.cn.

## 1 Introduction

The exploration of superconductors is of great interest in condensed matter physics and materials science because of their unique potential value in civil and industrial application[1-4]. Particularly, alkali metals as electrides[5-7] exhibit superconductivity which is quite different from hydrogen-rich superconductors[8, 9]. In electrides, the highly-localized electrons in the interstices of the lattice serve as anions, which are known as interstitial quasi-atom (ISQ)[10] and lead to strong core-ISQ polarization, resulting in exotic phenomena such as LA-TA splitting[11], universal surface interface metallic states[12], and colossal charge state of impurity atoms[7]. The presence of ISQs also can induce electronic phase transitions from metal to semiconductor, even to insulator, and with possible superconducting phase in them[13-17].

Recent high-pressure studies indicated that some Li-rich electrides might exhibit superconductivity generally, including $Li_5C$[18], $Li_5N$[19], $Li_6P$[20], $Li_9Te$[21], and $Li_8As$[22], in which it seems ISQs enhance superconductivity. However, in electrides $Li_6C$[23] and $Li_{10}Se$[24], the high superconducting critical temperature is attributed to the strong electron-phonon coupling derived from the synergy of interatomic coupling effect, while ISQs impede superconductivity. On the other hand, ISQs also contribute to the fusion and stability of intermetallic electride superconductors such as $Be_2Li$[25], LiCa[26], $Li_6Al$[27], $Li_8Cs$[28]. Recently, in the design of electrides $Li_8Au$[29] and $Li_8H_4$[30], the ISQs are adopted as a feature to achieve diverse topology to improve superconducting temperatures. Overall, the strong charge polarization in electrides provides a versatile tool to tune the electronic structure and superconductivity[22]. However, the role played by ISQs in superconductivity remains contentious. To establish a clear relation between ISQs and superconductivity is still imperative for to realize high-temperature electride superconductors.

Due to the localization of electron to the interstitial sites to achieve partial independence (i.e., only weakly associated with the nucleus), all insulating and semiconducting electrides are predicted having universal and robust metallic surface or interface states[12], even though some of them might be topologically trivial in terms of conventional topological band structure theory[31], which indicates electride might have unknown de facto real-space topology that beyond the conventional theories. Nevertheless, some electrides are indeed topological nontrivial in the electronic band structure[32, 33] according to conventional topological band theory. For instance, $Ca_3Pb$ possesses a topologically nontrivial band structure[34]. In $Cs_3O$ and $Ba_5N$, the anionic electrons induce an *s-p* band inversion and result in a nontrivial band topology[35]. Two dimensional electrides $Y_2C$, $Sr_2Bi$, and HfBr have also been reported being topological semimetals[36, 37]. Furthermore, the electride phase of simple metals also display nontrivial semimetallic electronic structure under pressure[38]. The high-pressure polymorphs of lithium are found to be topological nodal loop semimetals[39]; the alkali earth metals (Be, Mg, Ca, and Sr) exhibit Dirac nodal line and topological surface states[40].

In principle, the superconducting and conventional band topological states could be envisioned to emerge in the same electride. Their interplay might lead to interesting properties, such as the interaction between superconducting surface/edge current and colossal charge states[7] or the stationary longitudinal acoustic phonon mode[11, 12]. The search for novel topological superconducting electride thus is one of the major focuses in material science. As the lightest metallic element, lithium has a rich phase diagram under pressure, and exhibits abundant physicochemical properties[38, 41, 42]. In addition, its melting behaves like hydrogen (for example, the "U" shape in the melting curve after the melting temperature reaches a maximum[43]) under compression; but it maintains electride states that favor superconductivity before entering an semiconducting phase.

In this work, we report seven stable compounds with a wide stoichiometries spaning from 7:1 to 1:10 in Mg-Li system up to 500 GPa. Particularly, *Pm*-3*m* MgLi and *Pnma* MgLi become superconducting electrides with a noteworthy $T_c$ of 22.80 K and 8.91 K at high pressure, respectively. By composing and analyzing the variation of superconductivity in typical Li-contained electride, we obtain a practical relationship between the localization degree of ISQs and $T_c$, which could be employed as a guide to tune the superconducting temperatures by adjusting the localization of ISQs. Furthermore, it is demonstrated that *Pm*-3*m* MgLi and *Pnma* MgLi are also topological materials with a nontrivial $Z_2$ invariant and gapless surface states, thus as the first example of hosting superconductivity and topological nontrivial states in the same electride at the same time, which provides a platform for exploring the interplay between superconducting, band topological states, and non-nuclear centering of electrons.

## 2 Computational details

The thermodynamically stable structures of Mg-Li compounds under pressure are explored by using the ab initio evolutionary algorithm USPEX[44-46]. Structural optimization and electronic property calculations are performed by using the density functional theory (DFT)[47, 48] with the projector augmented wave (PAW)[49] method and the Perdew-Burke-Ernzerhof (PBE) parameterization of the generalized gradient approximation (GGA)[50] functional as implemented in the Vienna Ab-initio Simulation Package (VASP)[51, 52]. The energy cutoff for the plane wave basis set is chosen as 700 eV in all calculations. The largest spacing between k-points is set as $2\pi \times 0.015$ Å$^{-1}$ to generate automatic k-point meshes for the Brillouin zone sampling. All calculations are guaranteed to converge to $10^{-5}$ eV/atom for energy in the self-consistent field iterations and 0.01 eV/Å for the Hellmann-Feynman forces on atoms, respectively. The $1s^22s^1$ and $2s^22p^63s^2$ electrons are treated in the valence space for Li and Mg, respectively. The Bader charge analysis[53] is employed to characterize the charge state of atoms and the charge transfer between atoms and ISQs. Lattice dynamics are calculated by using the small displacement and supercell approach as implemented in our homemade MyPHON code[54], as well as the open source PHONOPY[55] code. The formation enthalpy ($\Delta H$) of $Mg_mLi_n$ with respect to the elemental Mg and Li at each pressure is defined as:

$$\Delta H(\mathrm{Mg}_m\mathrm{Li}_n) = [H(\mathrm{Mg}_m\mathrm{Li}_n) - mH(\mathrm{Mg}) - nH(\mathrm{Li})]/(m+n) \quad (1)$$

where $H$ is the enthalpy of the studied compound structure or the most stable elemental reference structure at the given pressure, respectively.

The electron-phonon coupling (EPC) and superconductivity are computed using density functional perturbation theory (DFPT), as implemented in the QUANTUM ESPRESSO (QE) code[56]. The ultrasoft pseudopotential for Li and Mg with a kinetic energy cutoff of 80 Ry are employed. Self-consistent electron density and EPC are calculated by employing carefully checked different $k$-meshes and $q$-meshes: 24×24×24 $k$-meshes and 12×12×12 $q$-meshes for *Pm*-3*m* MgLi, 20×20×20 $k$-meshes and 5×5×5 $q$-meshes for *Pnma* MgLi, 16×16×6 $k$-meshes and 8×8×3 $q$-meshes for *I*4/*mmm* $Mg_2Li$, respectively. The superconducting critical temperature $T_c$ is estimated using the Allen–Dynes-modified McMillan formula[57, 58] with correction given as follows:

$$T_c = \frac{\omega_{log}}{1.2} exp\left[-\frac{1.04(1+\lambda)}{\lambda-\mu^*(1+0.62\lambda)}\right] \quad \lambda \leq 1.5 \quad (2)$$

$$T_c = \frac{f_1 f_2 \omega_{log}}{1.2} exp\left[-\frac{1.04(1+\lambda)}{\lambda-\mu^*(1+0.62\lambda)}\right] \quad \lambda > 1.5 \quad (3)$$

where the effective Coulomb repulsion is taken as $\mu^* = 0.1$. The EPC parameter $\lambda$, the logarithmic average frequency $\omega_{log}$, the strong coupling and shape correction factors $f_1$ and $f_2$ are:

$$\lambda = 2\int \frac{\alpha^2 F(\omega)}{\omega} d\omega \quad (4)$$

$$\omega_{log} = \frac{2}{\lambda}\int \frac{\alpha^2 F(\omega)\ln\omega}{\omega} d\omega \quad (5)$$

$$f_1 = \sqrt[3]{1+\left[\frac{\lambda}{2.46(1+3.8\mu^*)}\right]^{3/2}} \quad (6)$$

$$f_2 = 1 + \frac{\left(\frac{\varpi_2}{\omega_{log}}-1\right)\lambda^2}{\lambda^2+\left[1.82(1+6.3\mu^*)\left(\frac{\varpi_2}{\omega_{log}}\right)\right]^2} \quad (7)$$

$$\varpi_2 = \sqrt{\frac{2}{\lambda}\int \omega\alpha^2 F(\omega) d\omega} \quad (8)$$

Furthermore, using Wannier representations, we also calculate the surface states of the studied compounds using the iterative Green's function as provided in the WANNIERTOOLS package[59, 60].

## 3 Results and discussion

The phase stability of $Mg_mLi_n$ system is investigated by calculating their enthalpy of formation up to 500 GPa. The compounds located on the convex hull are thermodynamically stable against any type of decomposition. As shown in Fig.1 (a), the predicted convex hulls suggest that Mg-rich compounds are easier to form in a wide range of pressures, while the only Li-rich compound $MgLi_{10}$ with space group *P*-1 is stable up to 118 GPa. Figure 1(b) shows the phase diagram and crystal structure of the stable phases of $Mg_mLi_n$. The ground-state phase of MgLi is found to be in the CsCl-type[61] structure (space group *Pm*-3*m*, Pearson symbol cP2) within a pressure range of 0-213 GPa. At higher pressures, MgLi decomposes into $Mg_2Li$ and Li element, as shown in Fig.1(a), Fig.S1(b), and Fig.S2. Interestingly, MgLi becomes the ground state again and is stabilized into the orthorhombic MnP-type[62] structure (oP8, space group *Pnma*) when between 350 GPa and 500 GPa.

For Mg-rich compositions, $Mg_3Li_2$ of $Os_2Al_3$-type[63] and $Mg_2Li$ of $MoSi_2$-type[64] are stable up to 174 GPa and 335 GPa, respectively. Both of them are ordered in a body-centered tetragonal (bct) lattice with space group *I*4/*mmm*. $Mg_3Li$ crystallizes into the $Na_3Cl$-type[65] structure with space group *P*4/*mmm* and maintains its stability between 4 GPa and 414 GPa. At elevated pressures, the bct-$Mg_4Li$ becomes stable from 50 to 476 GPa. The structure of $Mg_4Li$ also derives from $Os_2Al_3$[63] and crystallizes in the same space group (*I*4/*mmm*). Their phase diagram at 0 K is summarized in Fig. 1(b).

Furthermore, we also identified several metastable phases based on the fact that their formation enthalpies are not on the convex hull but is very close to it and exhibits dynamical stability (i.e., without any imaginary frequency in the phonon spectrum, see Fig.S3). The corresponding phases are: *P*-3*m*1 $MgLi_2$ (170−500 GPa), *Pm*-3*m* MgLi (213−250 GPa), *Pnma* MgLi (315−350 GPa), *I*4/*mmm* $Mg_3Li_2$ (174−300 GPa), *I*4/*mmm* $Mg_2Li$ (335−410 GPa),

*P*4/*mmm* $Mg_3Li$ (0−4 GPa, 414-500 GPa), *I*4/*mmm* $Mg_4Li$ (0−50 GPa, 476−500 GPa), and *P*4/*mmm* $Mg_5Li$ (0−500 GPa).

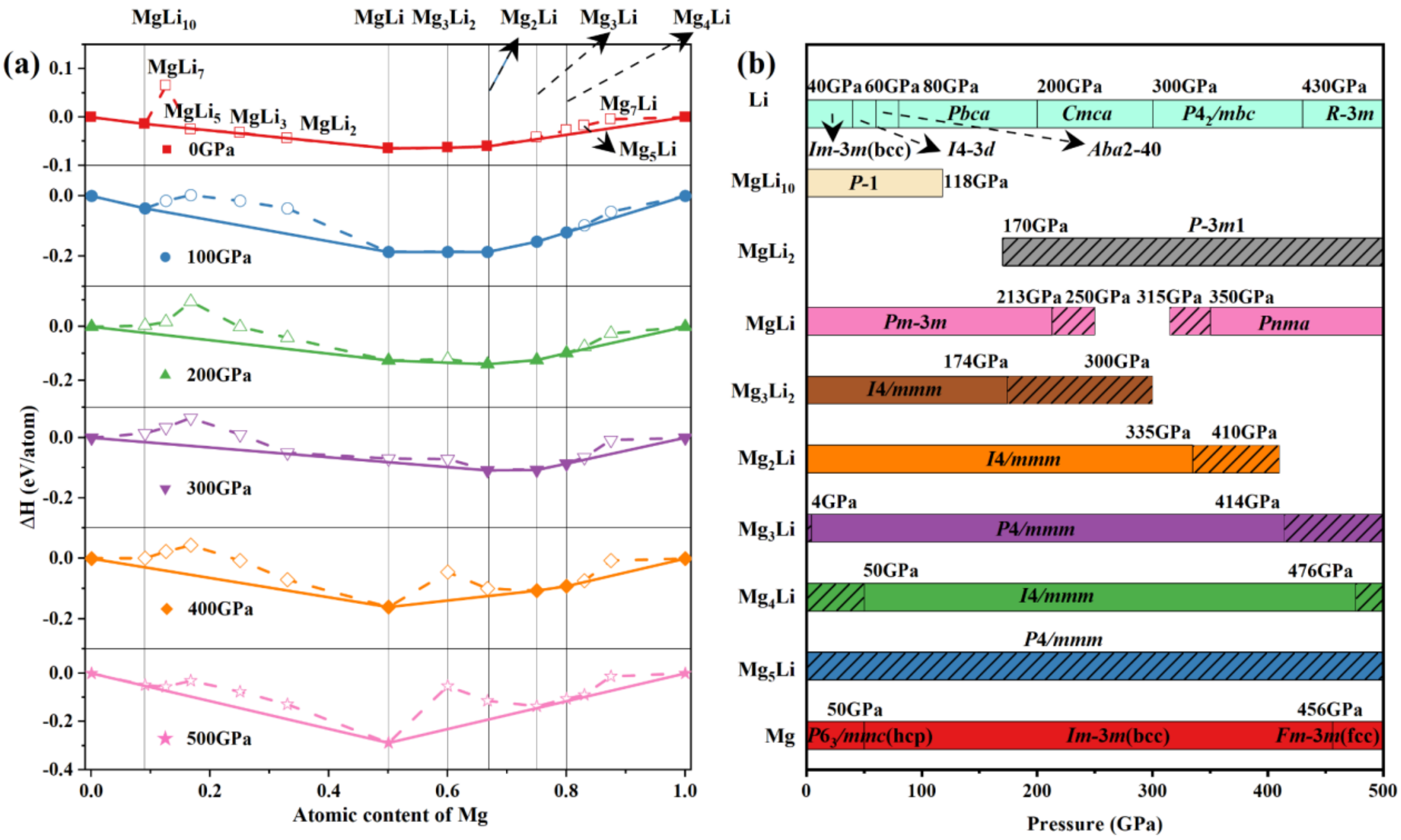

**Figure 1.** (Color online) Stability of $Mg_mLi_n$ compounds under pressure. (a) Enthalpy of formation of $Mg_mLi_n$ at selected pressures up to 500 GPa, in which the thermodynamically stable compounds are shown in solid symbols, while metastable ones are in open symbols; (b) Predicted phase diagram for stable $Mg_mLi_n$ compounds, where the black slashed regions indicate that the corresponding phases are metastable within the given pressure range. Though that the compounds of $MgLi_5$, $MgLi_3$, $MgLi_2$ at 0 GPa, $MgLi_2$ at 300 GPa, $Mg_5Li$ at 400 GPa, and $MgLi_{10}$, $Mg_3Li$, $Mg_4Li$, $Mg_5Li$ at 500 GPa are very close to the convex hull (within 3~9 meV/atom), they are denoted as metastable phases here.

As shown in Fig.2 (b), each Li atom in *Pm*-3*m* MgLi (B2 type structure) is surrounded by eight nearest Mg atoms and six second-nearest Li atoms. The corresponding Li–Li and Li−Mg distances (2.40 Å and 2.07 Å) are shorter than those in *Pm*-3*m* LiAl (2.54 Å and 2.20 Å)[66] at 200 GPa due to the stronger bonding in the former. In *Pnma*-MgLi, the Li atoms are bonded to six Mg atoms, with the Li-Mg bond distance of 1.64-1.73 Å at 500 GPa as shown in Fig. S4. On the other hand, the Mg-rich phase can be viewed as a host-guest structure, similar to Li-Al alloys[27]. The *I*4/*mmm* $Mg_3Li_2$ phase can be considered as the cubic lattice of Li-Mg inserted into the two-layer Mg cubic lattice as a guest. The Li atom is bonded in an eight-coordinated geometry of two inequivalent Mg sites. The first site is 4*e* (0.5, 0.5, 0.68865), and each Mg atom is bonded to eight equivalent Li atoms in a body-centered cubic geometry. The second site is 2*a* (0.0, 0.0, 0.0), and each Mg atom is bonded to four Li atoms. The corresponding Li-Mg distance at 100 GPa is 2.23 Å and 2.26 Å, respectively.

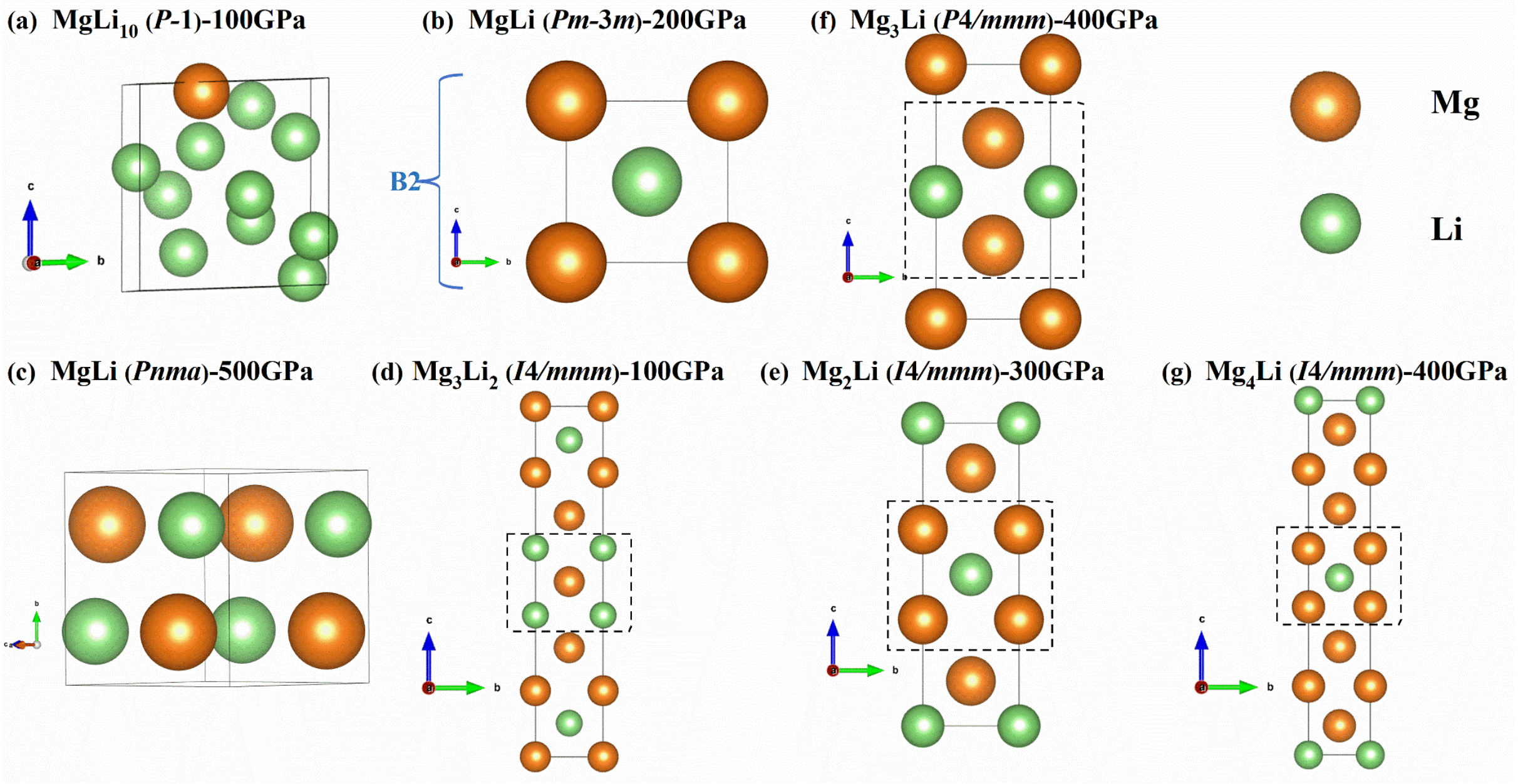

**Figure 2.** (Color online) Structures of stable $Mg_mLi_n$ compounds under high pressure: (a) *P*-1 $MgLi_{10}$ at 100 GPa; (b) *Pm*-3*m* MgLi at 200 GPa; (c) *Pnma* MgLi at 500 GPa; (d) *I*4/*mmm* $Mg_3Li_2$ at 100 GPa; (e) *I*4/*mmm* $Mg_2Li$ at 300 GPa; (f) *P*4/*mmm* $Mg_3Li$ at 400 GPa; (g) *I*4/*mmm* $Mg_4Li$ at 400 GPa. Orange and green spheres

represent Mg and Li atoms, respectively.

The structural features of $Mg_2Li$-*I*4*/mmm*, $Mg_3Li$-*P*4*/mmm*, and $Mg_4Li$-*I*4*/mmm* are similar, where each Li atom is coordinated with four Li atoms and eight Mg atoms. In $Mg_2Li$-*I*4*/mmm* phase, the shortest Li−Li and Li–Mg distance at 300 GPa is 2.31Å and 1.96 Å, respectively. In $Mg_3Li$-*P*4*/mmm* and $Mg_4Li$-*I*4*/mmm*, there are also two inequivalent Mg sites, with the Li–Li, Li–Mg1 and Li–Mg2 bond lengths of 2.22 Å, 1.89 Å, and 2.48 Å at 400 GPa, respectively. The detailed structural information of metastable $Mg_mLi_n$ compounds can be found in Fig.S5 and Table S1 of Supporting Information (SI).

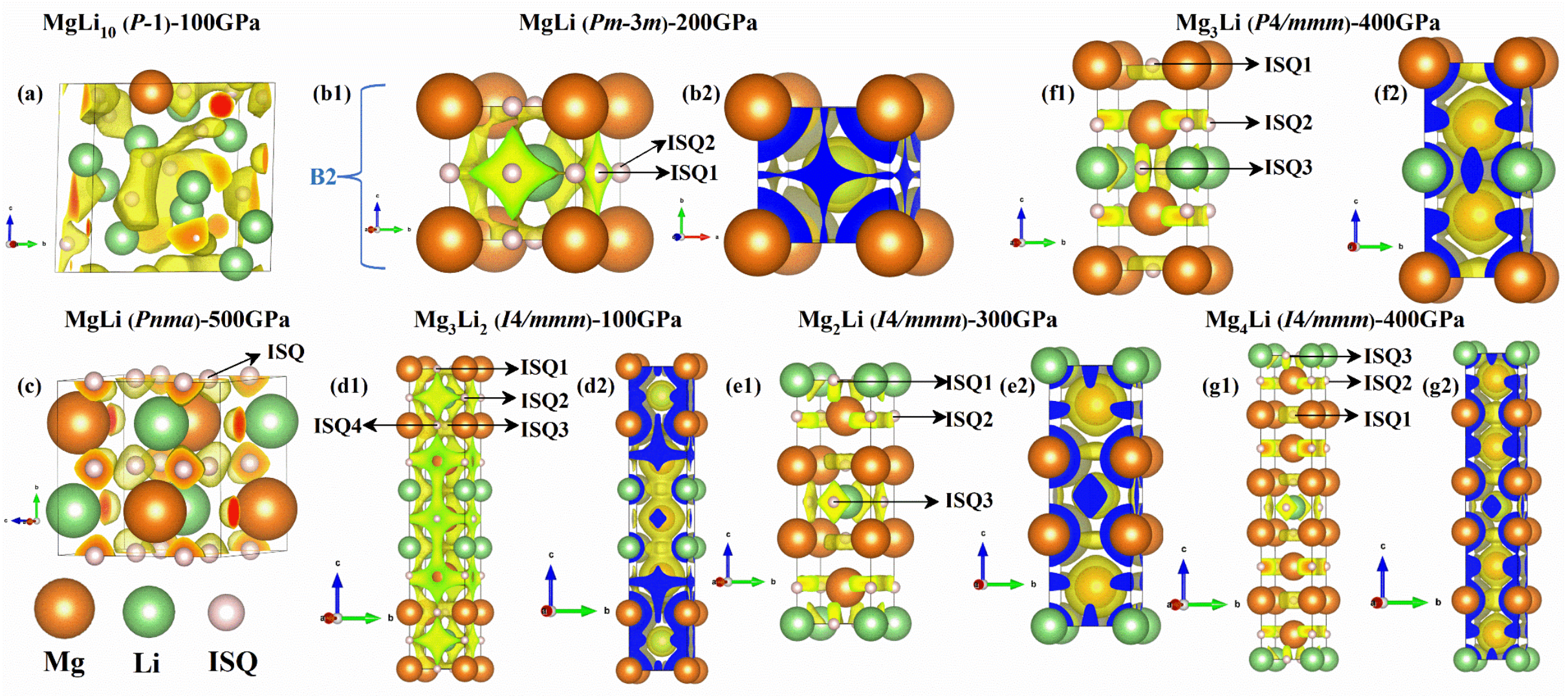

**Figure 3.** (Color online) Electron localization function (ELF) of $Mg_mLi_n$: (a) *P*-1 $MgLi_{10}$ at 100 GPa, (isosurface=0.75); (b1) *Pm*-3*m* MgLi at 200 GPa, (isosurface=0.60); (c) *Pnma* MgLi at 500 GPa, (isosurface=0.80); (d1) *I*4*/mmm* $Mg_3Li_2$ at 100 GPa, (isosurface=0.60); (e1) *I*4*/mmm* $Mg_2Li$ at 300 GPa, (isosurface=0.65); (f1) *P*4*/mmm* $Mg_3Li$ at 400 GPa, (isosurface=0.65); (g1) *I*4*/mmm* $Mg_4Li$ at 400 GPa, (isosurface=0.65). Charge density of $Mg_mLi_n$: (b2) *Pm*-3*m* MgLi at 200 GPa, (isosurface=0.0425 e/Bohr$^3$); (d2) *I*4*/mmm* $Mg_3Li_2$ at 100 GPa, (isosurface=0.0325 e/Bohr$^3$); (e2) *I*4*/mmm* $Mg_2Li$ at 300 GPa, (isosurface=0.052 e/Bohr$^3$); (f2) *P*4*/mmm* $Mg_3Li$ at 400 GPa, (isosurface=0.06 e/Bohr$^3$); (g2) *I*4*/mmm* $Mg_4Li$

at 400 GPa, (isosurface=0.06 e/Bohr$^3$). Orange, green and pink spheres represent Mg atoms, Li atoms and ISQ, respectively. Note that the structures in Fig.2 (b) and (d-g) are all derived from BCC lattice, their ELF as shown in Fig.3(b1) and (d1-g1) are similar to the BCC phase of lithium for the subunit that is of B2 type, and are similar to the uniaxially tensile BCC phase of lithium along *z* direction for those subunits with mixed occupation of atoms other than the B2 type.

The electron localization function (ELF) and charge density are calculated and shown in Fig.3. It is evident that all $Mg_mLi_n$ compounds show unambiguous localized interstitial electron, indicating they are indeed electrides according to the "ELF-charge" criterion advocated in Ref. [7]. Moreover, the calculated Bader charge (listed in Table S2 of SI) reveals that the valence states of both Li and Mg atoms are positive, confirming the charge transfer from atomic nuclei to ISQs. In the Li-rich phase ($MgLi_{10}$), the integrated number of electrons on Li is approximately -0.97 electrons, in other words, with a nominal oxidation state of +1. The localized electrons in interstices mainly come from Li atoms rather than from Mg, since Li has a smaller electronegativity than Mg at 100 GPa[67]. In *Pm*-3*m* MgLi, the anionic electrons are localized at two non-equivalent octahedral interstitial sites, forming ISQ1 and ISQ2 with a Bader charge of 0.44 e and 0.14 e, respectively. By contrast, the ISQs (with a charge state of 0.92 e) are arranged in layers in *Pnma* MgLi. The *I*4/*mmm* $Mg_3Li_2$ is derived from BCC lattice, with its ELF similar to the pristine or uniaxially tensile BCC phase of lithium, depending on whether the corresponding subunit is of the B2 type or not. The averaged Bader charge on the ISQs in this phase is: 0.55 e (ISQ1), 0.64 e (ISQ2), 0.68 e (ISQ3), 0.50 e (ISQ4), respectively. There are three inequivalent interstitial sites in *I*4/*mmm* $Mg_2Li$, leading to ISQ1 with 0.64 e,

ISQ2 with 0.84 e, and ISQ3 with 0.64 e, respectively. Excess electrons in both of them are localized in the vertex-sharing octahedral interstitials. The Bader charge analysis of $Mg_3Li$-*P*4/*mmm* gives 1.11 e (ISQ1), 0.89 e (ISQ2), 0.61 e (ISQ3); and 1.15 e (ISQ1), 0.85 e (ISQ2), 0.61 e (ISQ3) for $Mg_4Li$-*I*4/*mmm*, respectively.

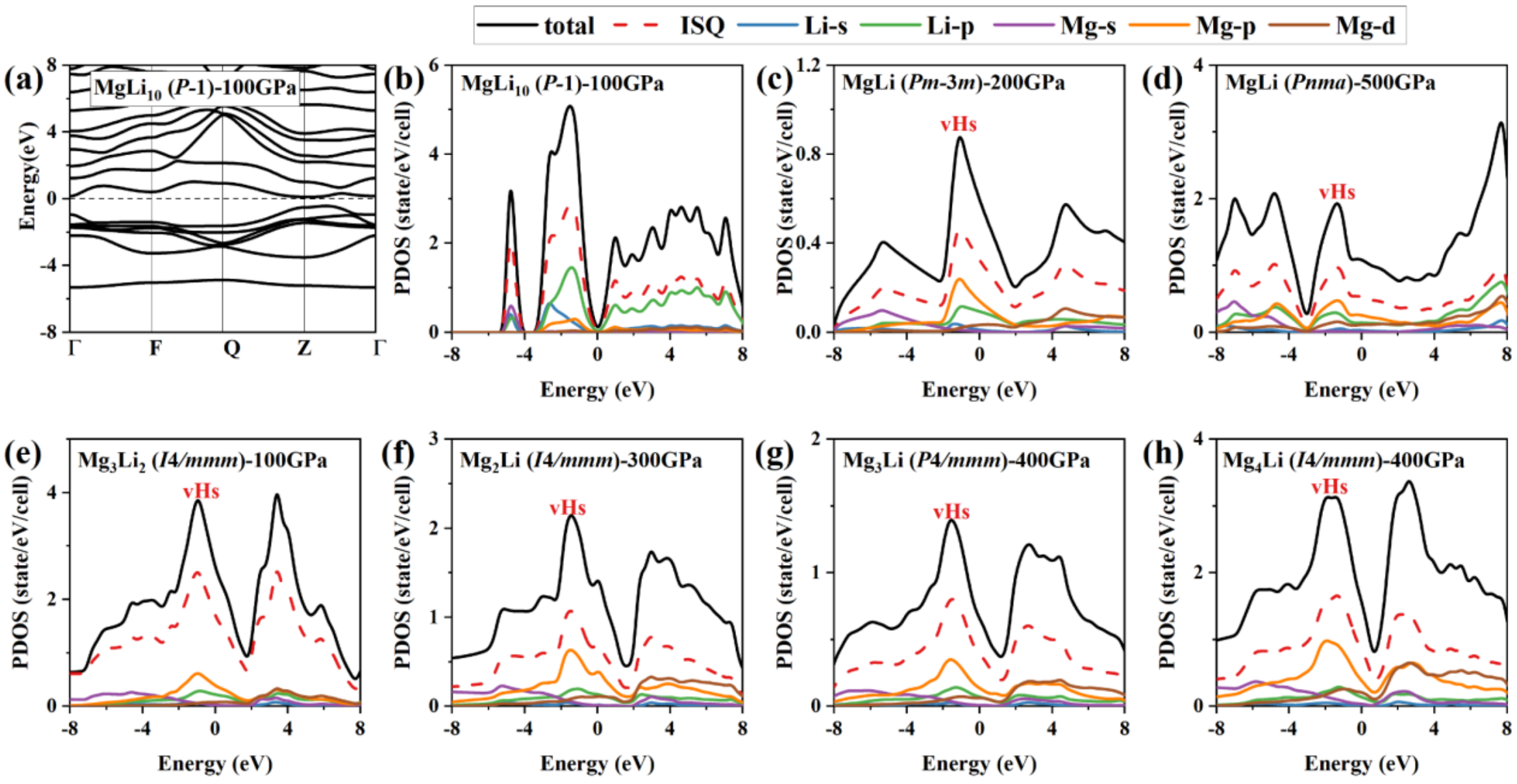

**Figure 4.** (Color online) (a) Band structure of *P*-1 $MgLi_{10}$ at 100 GPa; PDOS for (b) *P*-1 $MgLi_{10}$ at 100 GPa, (c) *Pm*-3*m* MgLi at 200 GPa, (d) *Pnma* MgLi at 500 GPa, (e) *I*4/*mmm* $Mg_3Li_2$ at 100 GPa, (f) *I*4/*mmm* $Mg_2Li$ at 300 GPa, (g) *P*4/*mmm* $Mg_3Li$ at 400 GPa, (h) *I*4/*mmm* $Mg_4Li$ at 400 GPa, respectively.

Electrides can exhibit interesting electronic properties, such as the semiconducting phases in compressed Li (*C*2[68] and *Aba*2[69]), Na (hP4[70] and oP8[71]), Li-Na compounds[72], $Ca_2N$-II, $Sr_2N$-II, and $Ba_2N$-IV[73]. We thus explored the band structure and projected density of states (PDOS) of the discovered Mg-Li electrides, as shown in Fig. 4. The electride $MgLi_{10}$ is discovered as a semiconductor with an energy gap of 0.08 eV at 60GPa, which increases to 0.22 eV at 100 GPa, as calculated by the revised Heyd-Scuseria-Ernzerhof screened hybrid functional (HSE06). As shown in Fig.4(b)-(h), the dominant contribution for the DOS at the Fermi level ($E_F$) comes from the interstitial electron. Moreover, there is not only a strong hybridization between ISQ and the Li-2*p*, Mg-2*p*, and Mg-3*d* electrons, but also remarkable van Hove singularities (vHs) dominated by the anionic electrons near the Fermi level. The increased DOS at the $E_F$ of MgLi and $Mg_2Li$ suggest the possibility of the emergence of superconductivity in such componds[74].

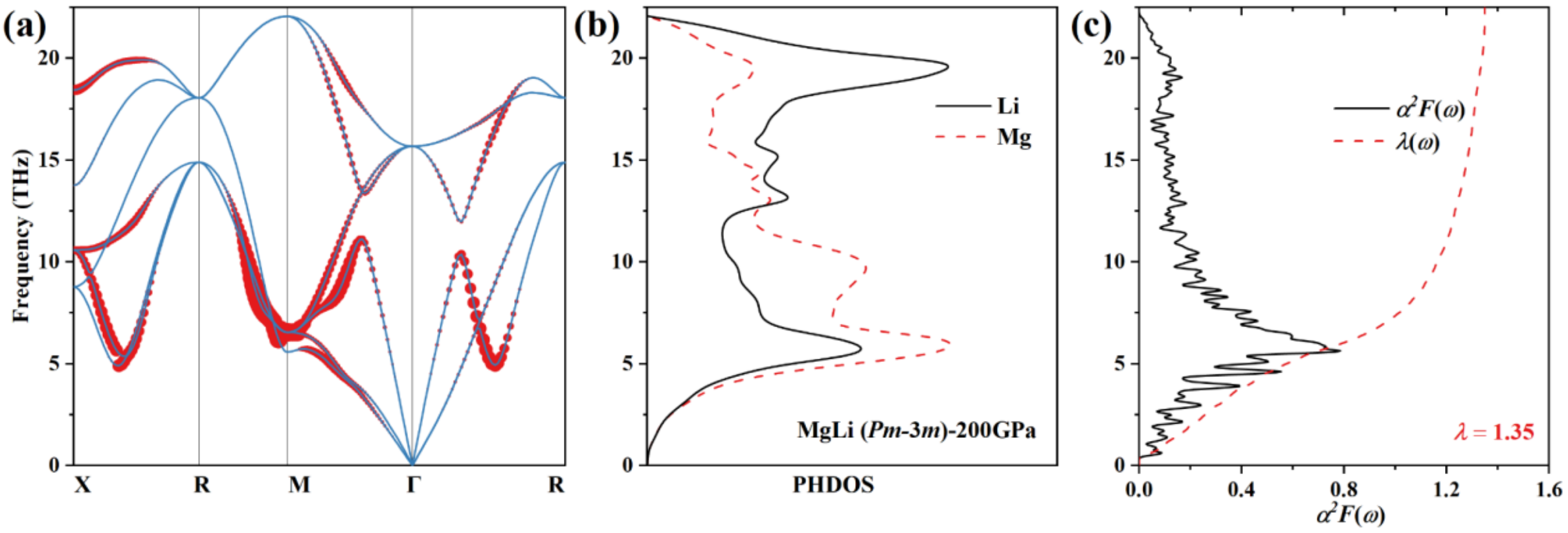

**Figure 5.** (Color online) (a) Phonon dispersion spectrum, (b) projected phonon density of states (PHDOS), (c) Eliashberg spectral function $\alpha^2F(\omega)$ and electron-phonon coupling constant $\lambda$ of *Pm*-3*m* MgLi at 200 GPa, respectively. The size of the solid red circles in (a) is proportional to the electron phonon coupling strength.

To investigate the phonon-mediated superconductivity in Mg-Li alloys, we calculated their phonon dispersion spectra and electron-phonon coupling (EPC) constants $\lambda$, as depicted in Figs. 5, S6, S7, and Table 1. Using a Coulomb pseudopotential of $\mu^*= 0.1$, the calculated EPC constant $\lambda$ of the *Pm*-3*m* MgLi is around 1.35 at 200 GPa (Fig.5(c)), and the superconductivity critical temperature $T_c$ is estimated to be around 22.80 K. It is worth noting that the main contribution to the superconductivity comes from the low-frequency phonon modes along the R-M-Γ directions (see Fig.5(a), especially at the M point). From the PHDOS and Eliashberg spectral function, one can determine that the low-frequency region below 13 THz contribute 78.6% to the EPC constant due to the hybridization of the vibrations of Li with Mg atom. The modes at higher frequency contribute 21.4% to the total $\lambda$, where the vibration of Li plays a critical role. As for *Pnma* MgLi and *I4/mmm* $Mg_2Li$, the EPC is weaker, with $\lambda$=0.79 and $\lambda$=0.45, resulting in a $T_c$=8.91 K and $T_c$=3.24 K, respectively. The low-frequency modes of the phonon spectrum also dominate in their total EPC.

**Table 1**. The calculated total EPC parameter $\lambda$, logarithmic average phonon frequency $\omega_{log}$, and superconducting critical temperature $T_c$ of *Pm*-3*m* MgLi at 200 GPa, *Pnma* MgLi at 500 GPa, and *I4/mmm* $Mg_2Li$ at 300 GPa, respectively.

| Phase | Pressure (GPa) | $\lambda$ | $\omega_{log}$ | $T_c$ (K) |
|---|---|---|---|---|
| *Pm*-3*m* MgLi | 200 GPa | 1.35 | 174.201 | 22.80 |
| *Pnma* MgLi | 500 GPa | 0.79 | 176.408 | 8.91 |
| *I4/mmm* $Mg_2Li$ | 300 GPa | 0.45 | 254.768 | 3.24 |

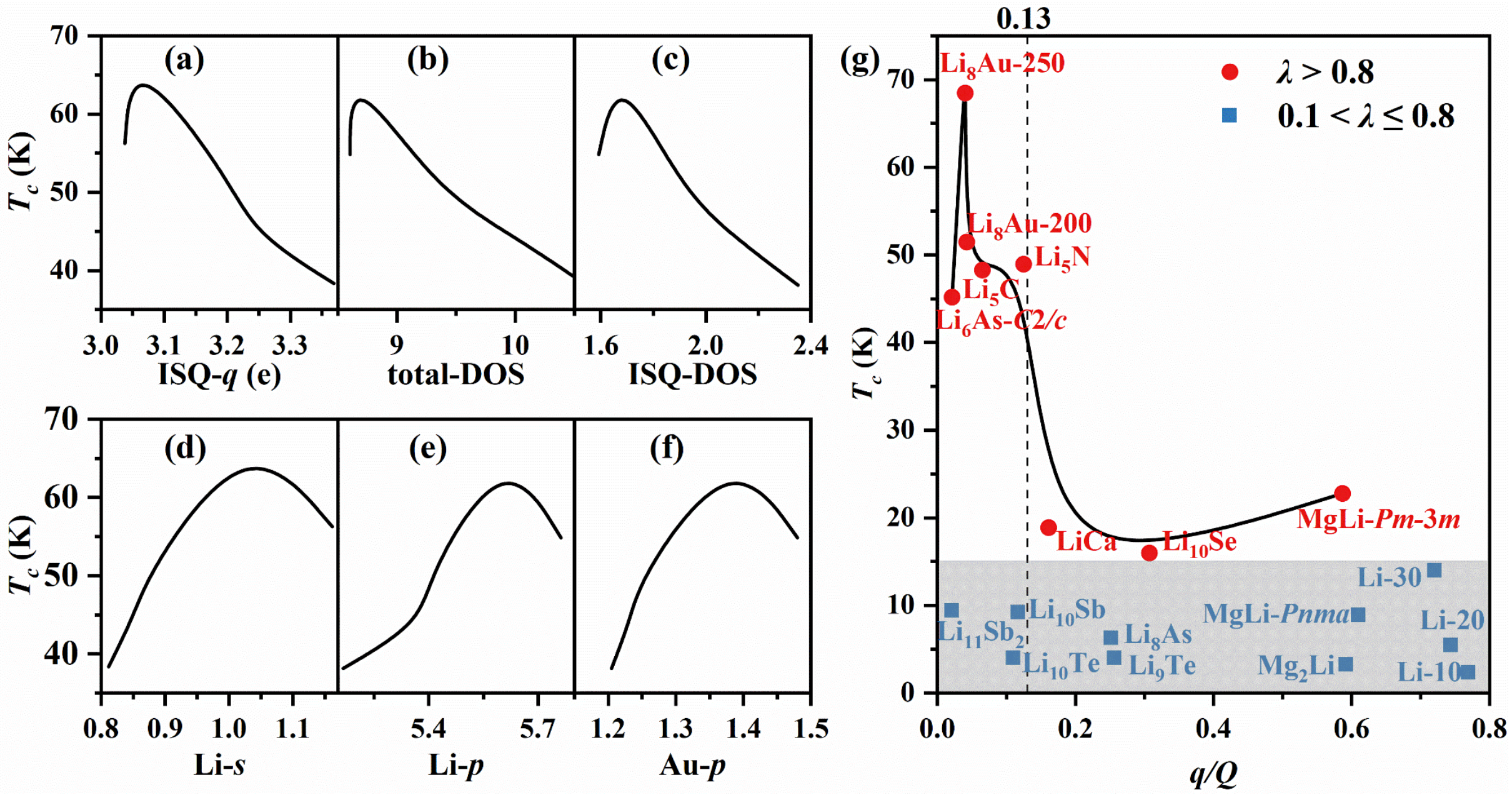

**Figure 6.** (Color online) The evolution of $T_c$ with (a) the charge of ISQs, (b) the total density of states (DOS)

at the Fermi level, (c) the DOS of ISQs at the Fermi level, (d) the DOS of Li-*s* orbital at the Fermi level, (e) the DOS of Li-*p* orbital at the Fermi level, (f) the DOS of Au-*p* orbital at the Fermi level in $Li_8Au$, respectively; (g) The evolution of $T_c$ with localization intensity $q/Q$ in typical Li-based electrides. The bule squares represent weakly coupled electrides ($0.1<\lambda \leqslant 0.8$), red spheres represent moderately coupled electrides ($\lambda>0.8$).

As mentioned above, Li-rich compounds exhibit general superconductivity. They in fact can be viewed as variants derived from the elemental Li that is also a superconducting electride. They should have a universal rule that governs the electron localization and superconductivity. To explore this, we took $Li_8Au$ as an example to illustrate the relationship between ISQs and superconductivity, which has the highest $T_c$ reported to date for all electrides. There are two stages as unveiled in Fig. 6(a)-(f): in the first stage ($T_c$: 55 K→65K), the ISQs enhance the density of states (DOS) at the Fermi level, and the $T_c$ values are positively correlated with the ISQs; In the second stage ($T_c$: 65 K→40K), the over-localized ISQs start to weaken the electron-phonon coupling, which lowers down the superconducting temperature. The two competing mechanisms result in a maximum critical point which bridges the free-electron gas model of metals to highly localized electrides.

In order to quantify this, let us define the ratio of Bader charge of ISQs ($q$) and the valence electron ($Q$) as the localization intensity, and divide the typical electride superconductors into two groups based on EPC constants $\lambda$: weakly coupled materials ($0.1<\lambda\leq 0.8$) and moderately coupled materials ($\lambda>0.8$). The superconducting critical temperatures of the weakly coupled electrids are below 15K, for which the localization intensity of ISQs hardly affects the superconductivity [see the blue area in Fig. 6(g)]. On the other hand, for the moderately coupled electrides, the superconducting temperatures increase and then decrease notably and finally become flattened with increasing *q/Q*. As shown in Fig. 6(g) and Table S3, the localization intensity of ISQs can be used to judge whether the superconducting temperature reaches the optimal value or not. For moderately coupled electrides, the superconductivity can be tuned by adjusting the localization intensity of ISQs. For the Mg-Li system we studied, the *Pm*-3*m* MgLi is moderately coupled electride, in which ISQs are over-localized (*q/Q* = 0.59, see the Table S3 in the Supporting Information) thus has a relatively low $T_c$. In this regard, weakening the localization may increase the superconducting critical temperatures. On the other hand, *Pnma* MgLi and *I4/mmm* $Mg_2Li$ are weakly coupled electrides, the localization intensity of ISQs does not affect their $T_c$.

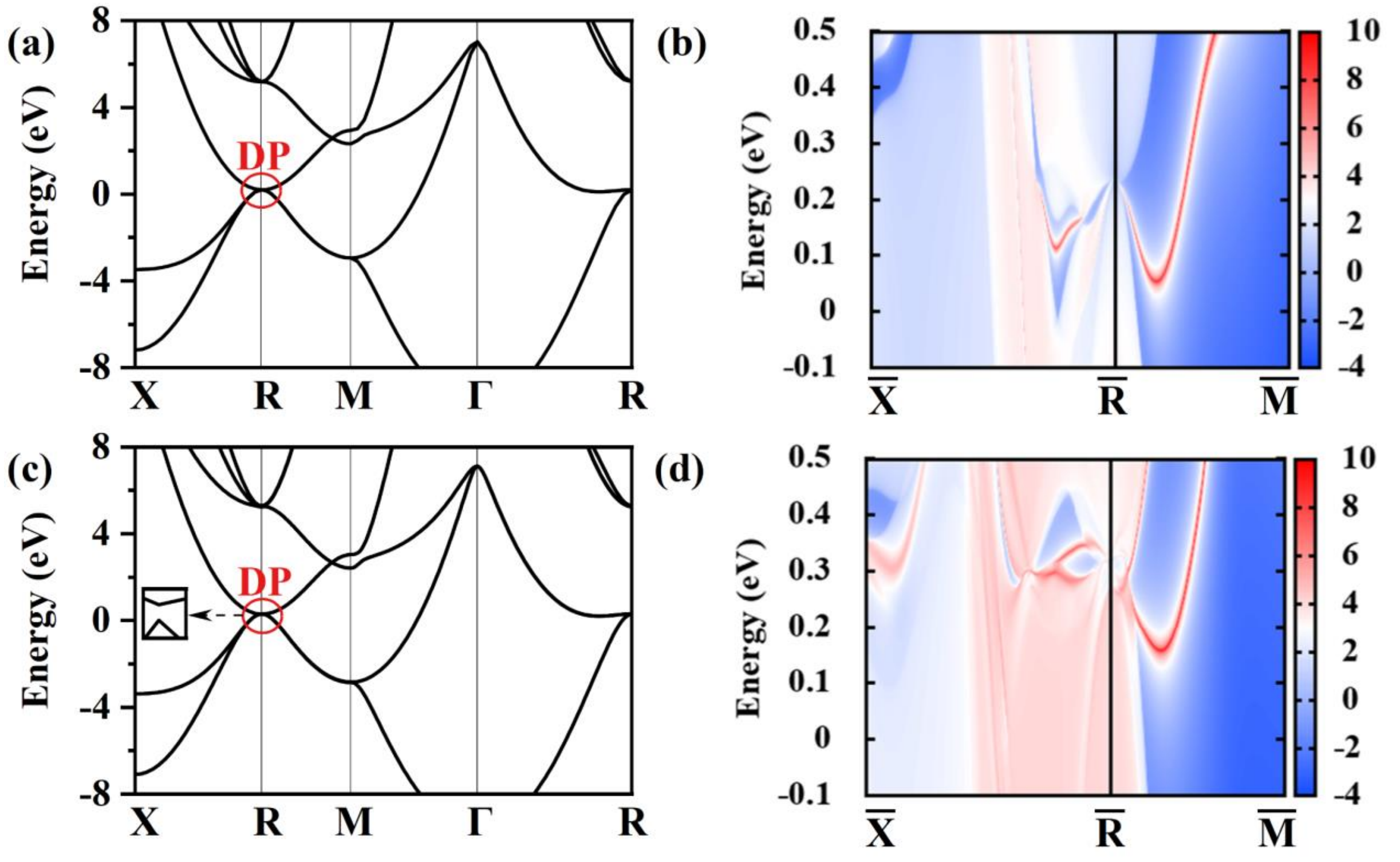

**Figure 7.** (Color online) (a) Bulk band structure and (b) (001) surface states of *Pm*-3*m* MgLi calculated without the SOC effec. (c) Bulk band structure and (d) (001) surface states of *Pm*-3*m* MgLi calculated with the SOC effec. Inset of (c) shows a zoom in of the DP with a small gap opened by SOC. The Fermi level is set to zero. The pressure is 200 GPa.

In the band structure of bulk *Pm*-3*m* MgLi, we found that the conduction and valence bands cross each other at the *R* point, leading to the formation of a band touching point (i.e., a Dirac point). Interestingly, the gapless Dirac point (DP) appears near the Fermi level if without spin-orbit coupling (SOC); whereas the DP becomes gapped when the SOC is turned on. Here, we have investigated its topological properties by studying the surface states (Fig. 7(b)) and the topologically invariant $Z_2$ indices (Table 2). The latter can be determined from the parity of the wave function at the time reversal invariant momentum (TRIM) points in the presence of time-reversal symmetry and inversion symmetry. The product of the parity for the occupied states at the TRIM points are calculated by $\delta(\boldsymbol{k_i}) = \prod_{N=1}^{14} \xi_{2N}(\boldsymbol{k_i})$, where $\xi_{2N}(\boldsymbol{k_i})$ represents the parity of the occupied valence bands at $\boldsymbol{k_i}$. The strong topological insulators are characterized by a $Z_2$ invariant $(v_0; v_1 v_2 v_3)$ = (1; 111). The gapless surface states can be identified near the Fermi level according to the bulk-boundary correspondence. The *Pm*-3*m* phase of MgLi electride thus could be described as a strong topological material, as evidenced by the calculated $Z_2$ = (1; 111) indices and the nontrivial surface states (Fig.7(d)). For the *Pnma* phase, the band-crossing point is located at the *S* point (see Fig. S8) and the gap is also opened up by SOC. The calculated $Z_2$ invariant (see Table S4 of SI) and metallic surface-states also confirm the nontrivial topological nature of MgLi-*Pnma*. These results indicate that the MgLi electrides (*Pm*-3*m* and *Pnma*) all exhibit topologically nontrivial characteristics. Since superconductivity, band topology and interstitial localization are all derived from the quantum behavior of electrons. *Pm*-3*m* MgLi and *Pnma* MgLi demonstrate the multi-quantum state realized in this system.

**Table 2.** Parity products at TRIMs and the $Z_2$ indices of *Pm*-3*m* MgLi at 200 GPa.

| TRIM | Parity products | $v$ | $Z_2$ |
|---|---|---|---|
| X | - | $v_1$=0, $v_1'$=1 | |
| 3R | + | $v_2$=0, $v_2'$=1 | |
| 3M | - | $v_3$=0, $v_3'$=1 | (1;111) |
| Γ | - | $v_0$=1 | |

Finally, it is worth mentioning that the long-range interaction between atoms and ISQs could lead to an anomalous LA-TA splitting in electride, analogous to the LO-TO splitting in ionic compounds[11]. For the isotropic *Pm*-3*m* MgLi, the magnitude of such splitting is ($\Delta\omega = \omega_{LA} - \omega_{TA}$) 34.5 cm$^{-1}$ at 200 GPa (see Fig. S10 in the Supporting Information), which might have an impact on the electrical or optical properties of the surface state of this class of materials[12].

## 4. Conclusion

In summary, we have extensively explored the ground-state structures and phase diagram of Mg-Li system up to 500 GPa by combining crystal structure prediction and first-principles calculations. Seven compounds have been found to be stable, including $MgLi_{10}$, *Pm*-3*m* MgLi, *Pnma* MgLi, $Mg_3Li_2$, $Mg_2Li$, $Mg_3Li$, and $Mg_4Li$, respectively. The Mg-Li system seems to favor Mg-rich compounds compared to the Li-rich side in a wide range of pressures. Remarkably, all high-pressure compounds are electrides with various morphologies of ISQs.

Calculations of the electronic properties reveal that $MgLi_{10}$ is a semiconductor with almost all of the 2*s* electrons of lithium having been transferred into the interstitial sites. Interestingly, our calculations show that *Pm*-3*m* MgLi is a superconductor with a critical temperature $T_c$ of 22.8 K at 200 GPa; and *Pnma* MgLi and $Mg_2Li$ also manifest superconductivity. We found in the moderately coupled electrides (such as *Pm*-3*m* MgLi), the superconducting temperatures increase firstly and then decrease notably and finally become flattened with increasing localization intensity of ISQs. By contrast, in the weakly coupled electrides (such as *Pnma* MgLi and $Mg_2Li$), the ISQs hardly affect the superconductivity.

In particular, the valance and conduction bands of MgLi compounds (*Pm*-3*m* and *Pnma*) cross each other at the DPs if without SOC; but the gap at the DPs opens up near the Fermi level when with SOC. Both the gapless surface states and the calculated $Z_2$ invariant clearly confirm that these two phases of MgLi are band topologically nontrivial. The *Pm*-3*m* MgLi electride exhibits both superconductivity and topological behaviors, which also has the highest superconducting $T_c$ among all reported band topological electrides to date. The simultaneous emergence of the superconductivity and the band topology in the same electride may lead to potential applications in electride-based quantum devices.

## Author contributions

Dan Wang: Investigation, Methodology, Writing - original draft, Writing - review & editing. Hong X. Song: Methodology, Writing - review & editing. Qi D. Hao: Methodology, Writing - review & editing. Guang F. Yang: Methodology, Writing - review & editing. Hao Wang: Methodology. Lei L. Zhang: Writing - review & editing. Xiang R. Chen: Supervision, Project administration. Hua Y. Geng: Idea conceiving, Project design, Writing, Reviewing, and Editing, Supervision, Project administration, Software.

## Conflicts of interest

The authors declare no competing interests.

## Data availability

All data are included in the manuscript and supporting information.

## Code availability

The VASP code used in this paper is commercial software provided by VASP Software GmbH. All others are open-source codes that can be obtained via internet.

## Supporting Information

The Enthalpy difference for different Ecut; The enthalpy of formation, calculated phonon spectra, and histogram of interatomic separations of $Mg_mLi_n$; The structure of *P*-3*m*1 $MgLi_2$ and *P*4/*mmm* $Mg_5Li$; The structure information of $Mg_mLi_n$; The valence state of Mg and Li atoms and the Bader charge of ISQs in $Mg_mLi_n$ compounds. The phonon dispersion relations, PHDOS, $\alpha^2F(\omega)$ and EPC parameter $\lambda$ of *Pnma* MgLi and *I*4/*mmm* $Mg_2Li$; The charge state of ISQs($q$), valence electrons($Q$), $q/Q$, EPC parameter $\lambda$, and superconducting critical temperature $T_c$ of typical electrides that contain Li; The topological behaviors of MgLi-*Pnma*; Band structures of *I*4/*mmm* $Mg_3Li_2$, *I*4/*mmm* $Mg_2Li$, *P*4/*mmm* $Mg_3Li$, and *I*4/*mmm* $Mg_4Li$; LA-TA splitting of *Pm*-3*m* MgLi.

## Acknowledgments

This work was supported by the National Key R&D Program of China under Grant No.

2021YFB3802300, the Foundation of National key Laboratory of shock wave and detonation physics (No. 2023JCJQLB05401), the National Natural Science Foundation of China under Grant No. 12074274, and the NSAF under Grant No. U1730248. Part of the computation was performed using the supercomputer at the Center for Computational Materials Science (CCMS) of the Institute for Materials Research (IMR) at Tohoku University, Japan.

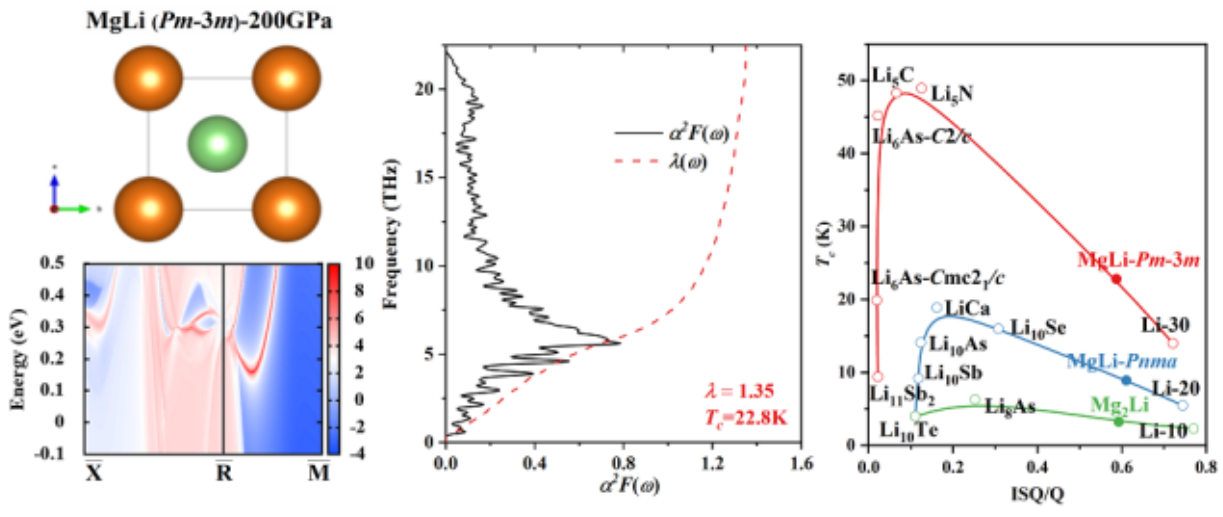

## TOC Graphic

# Supporting Information for

# Superconducting and Topological Properties in Mg−Li Electrides at High Pressures

Dan Wang,[1,2] Hong X. Song,[1] Qi D. Hao,[1,2] Guang F. Yang, [1] Hao Wang,[1] Lei L. Zhang,[3] Ying Chen,[4] Xiang R. Chen,[2*] Hua Y. Geng[1,5*]

[1] *National Key Laboratory of Shock Wave and Detonation Physics, Institute of Fluid Physics, China Academy of Engineering Physics, Mianyang, Sichuan 621900, P. R. China;*

[2] *Institute of Atomic and Molecular Physics, College of Physics, Sichuan University, Chengdu 610065, P. R. China;*

[3] *Institute of Nano-Structured Functional Materials, Huanghe Science and Technology College, Zhengzhou 450063, P. R. China;*

[4] *Fracture and Reliability Research Institute, School of Engineering, Tohoku University, Sendai 980-8579, Japan;*

[5] *HEDPS, Center for Applied Physics and Technology, and College of Engineering, Peking University, Beijing 100871, P. R. China.*

* To whom correspondence should be addressed. E-mail: s102genghy@caep.cn; xrchen@scu.edu.cn.

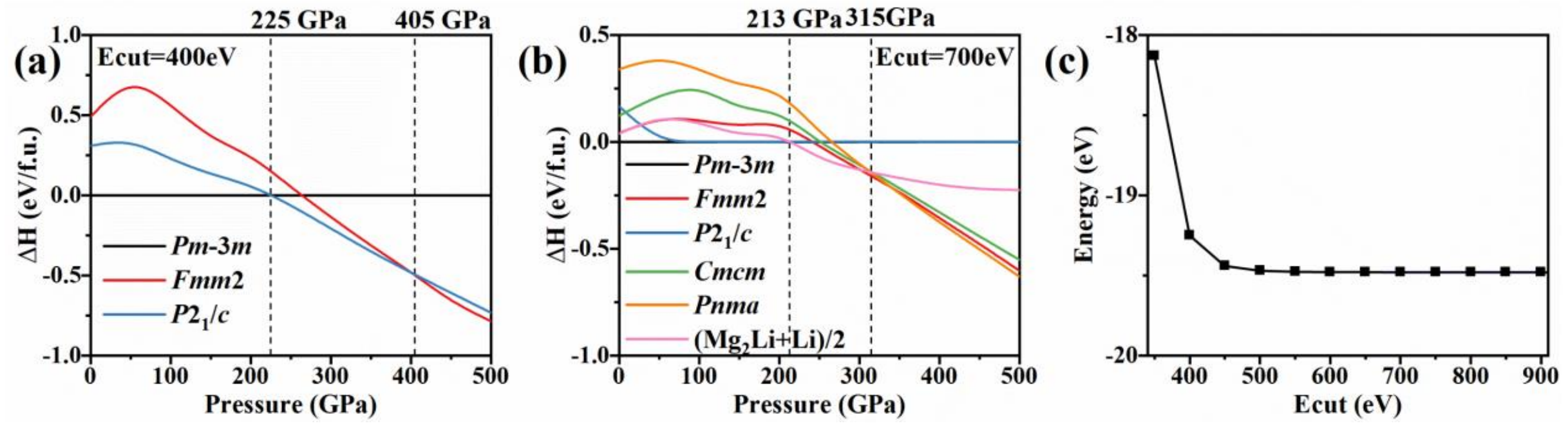

**Figure S1.** (Color online) (a) Enthalpy difference of MgLi in *Fmm*2 and *P*2$_1$/*c* phases with respect to the *Pm*-3*m* phase for Ecut: 400 eV; The dashed-lines indicate the pressure of the phase transition; (b) Enthalpy difference of MgLi in *Fmm*2, *P*2$_1$/*c*, *Cmcm*, and *Pnma* phases with respect to the *Pm*-3*m* phase for Ecut: 700 eV; (c) Convergence test for the energy cutoff (Ecut) of the plane wave basis.

As shown in Fig.S1, the *Pm*-3*m* MgLi transforms to the *P*2$_1$/*c* phase at 225 GPa and then transforms to the *Fmm*2 phase at 405 GPa if with a cut-off energy of 400 eV, which are consistent with the results of the previous theoretical study[1]. However, when the cut-off energy for the plane-wave basis is increased to 700 eV, the picture is changed: *Pm*-3*m* MgLi decomposes into $Mg_2Li$ and Li at 213 GPa and then becomes stable again into the *Pnma* phase at 315 GPa. Based on the convergence test for the energy cutoff of plane wave basis, the results of Ecut: 700 eV are more reliable.

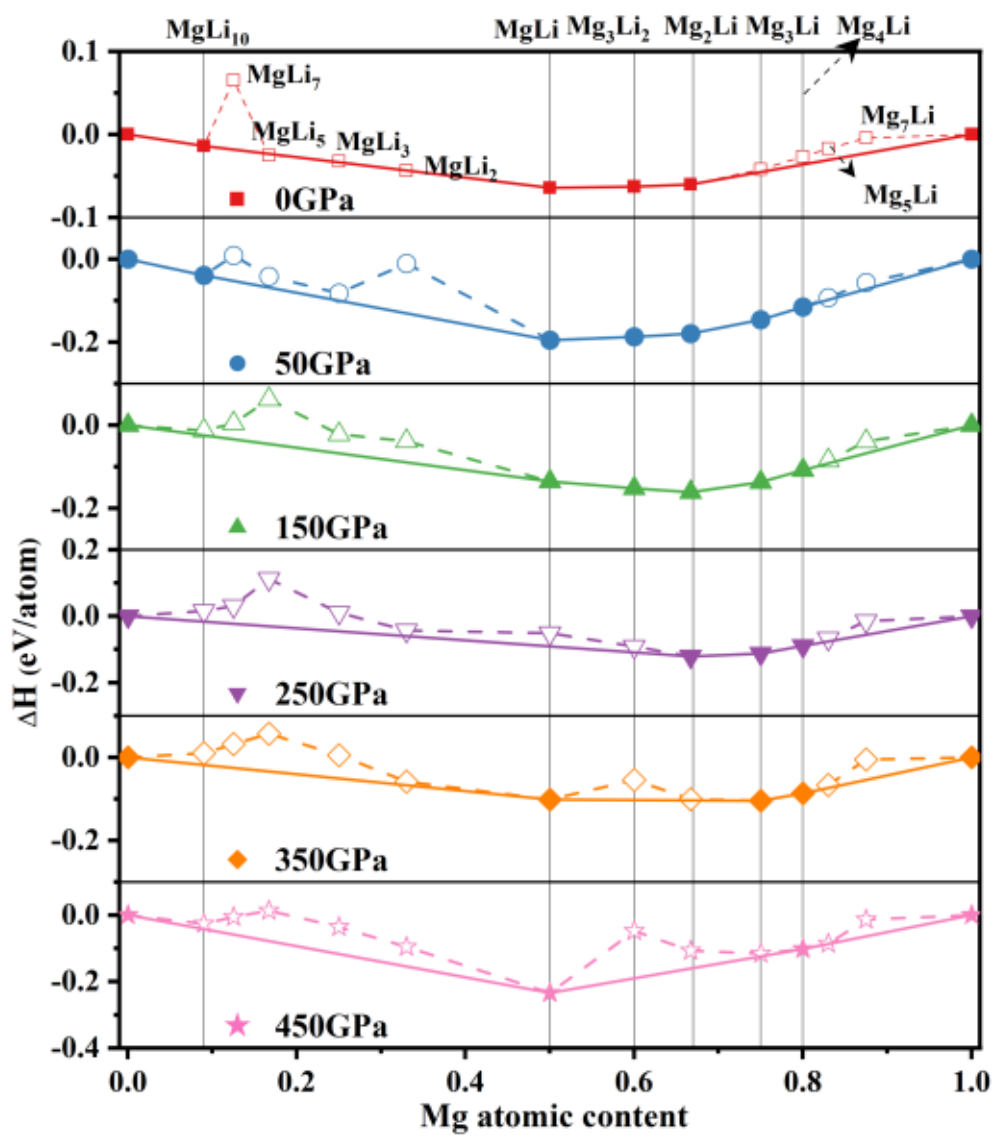

**Figure S2.** (Color online) Enthalpy of formation of $Mg_mLi_n$ under 0 GPa, 50 GPa, 150 GPa, 250 GPa, 350 GPa, and 450 GPa. The thermodynamically stable compounds are shown by solid symbols.

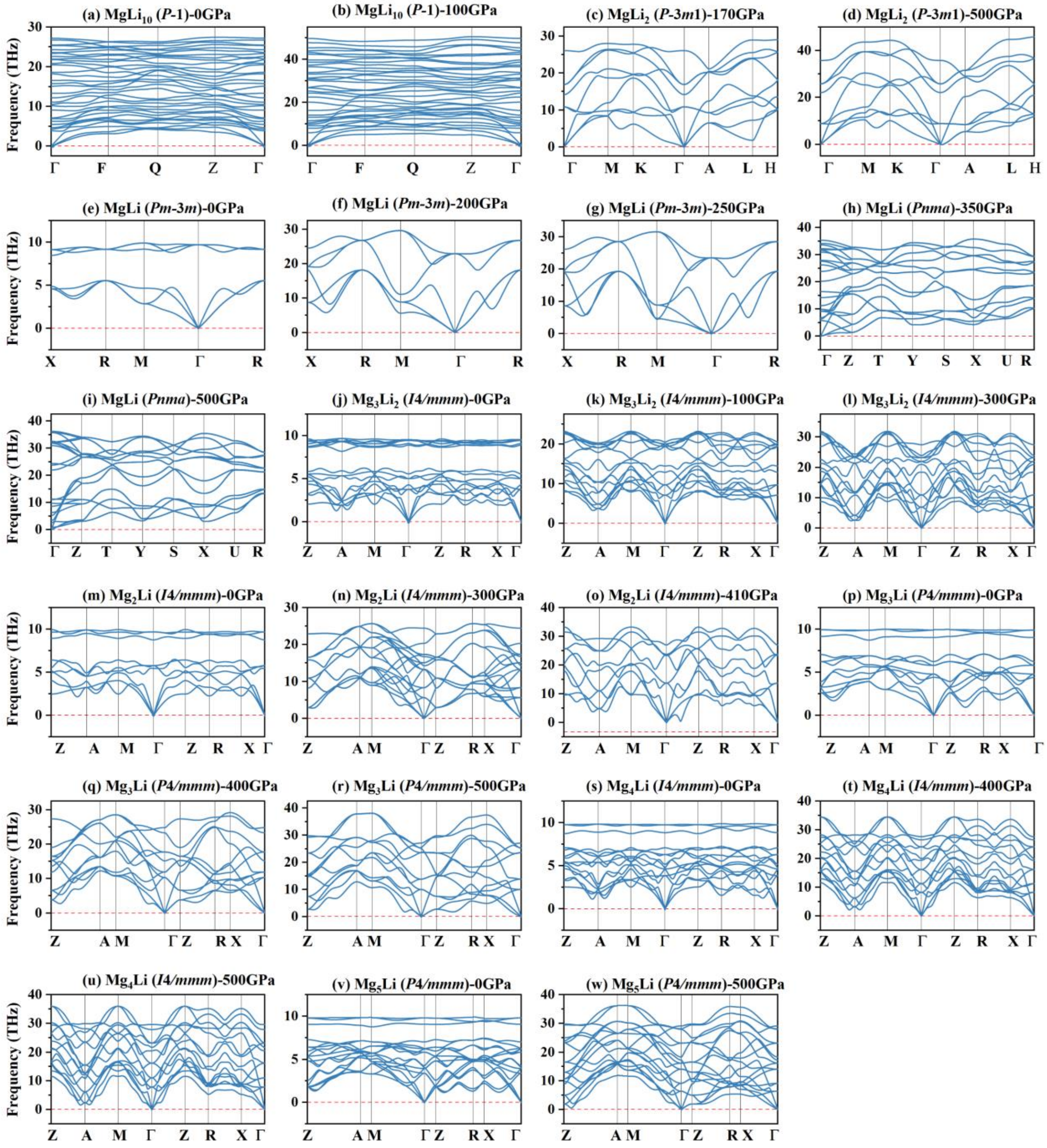

**Figure S3.** (Color online) Calculated phonon spectra of *P*-1 $MgLi_{10}$ at (a) 0 GPa and (b) 100 GPa; Calculated phonon spectra of *P*-3*m*1 $MgLi_2$ at (c) 170 GPa and (d) 500 GPa; Calculated phonon spectra of *Pm*-3*m* MgLi at (e) 0 GPa, (f) 200 GPa, and (g) 250 GPa; Calculated phonon spectra of *Pnma* MgLi at (h) 350 GPa, (i) 500 GPa; Calculated phonon spectra of *I*4/*mmm* $Mg_3Li_2$ at (j) 0 GPa, (k) 100 GPa, and (l) 300 GPa; Calculated

phonon spectra of *I4/mmm* $Mg_2Li$ at (m) 0 GPa, (n) 300 GPa, and (o) 410 GPa; Calculated phonon spectra of *P4/mmm* $Mg_3Li$ at (p) 0 GPa, (q) 400 GPa, and (r) 500 GPa; Calculated phonon spectra of *I4/mmm* $Mg_4Li$ at (s) 0 GPa, (t) 400 GPa, and (u) 500 GPa; Calculated phonon spectra of *P4/mmm* $Mg_5Li$ at (v) 0 GPa and (w) 500 GPa, respectively.

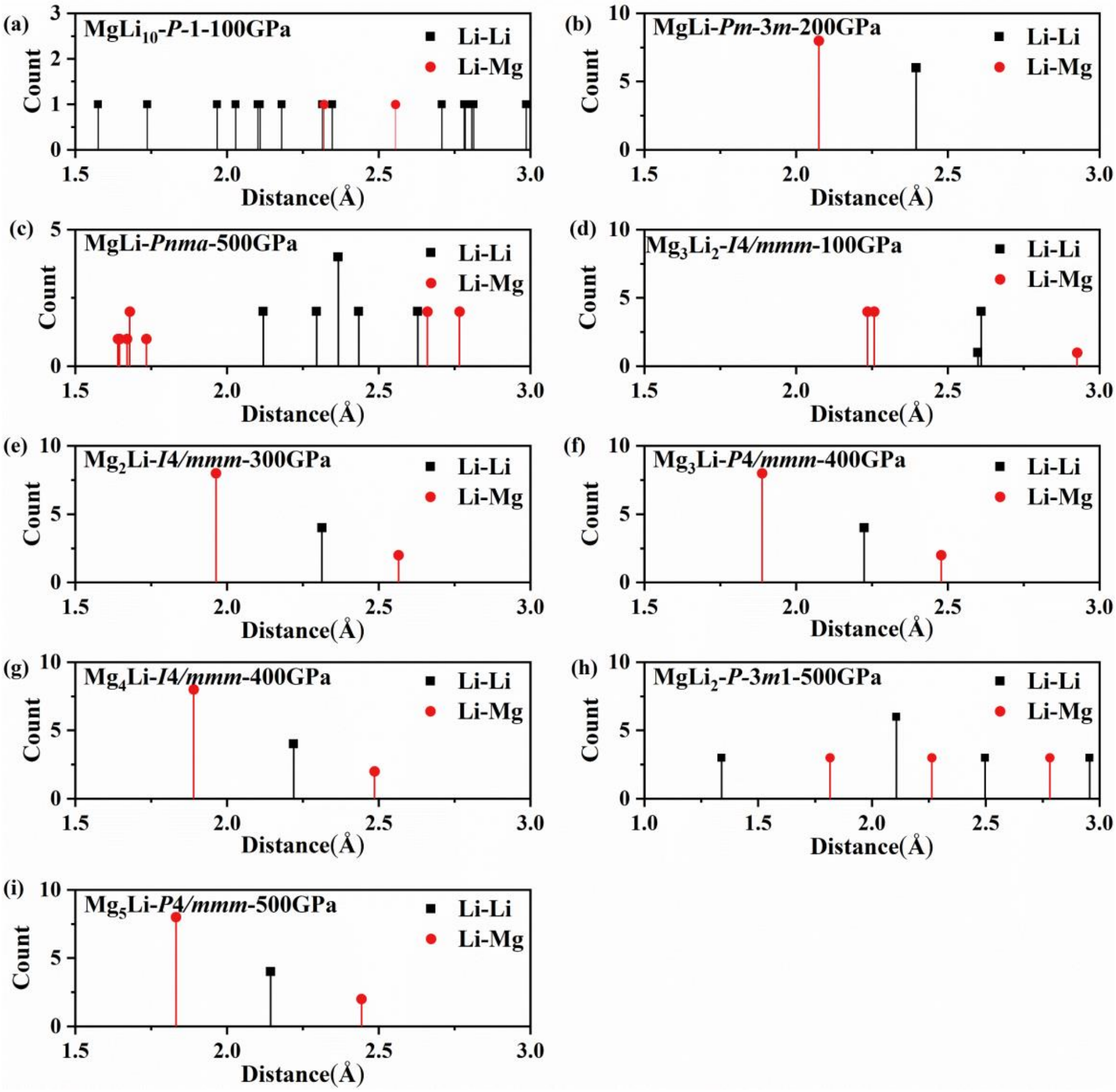

**Figure S4.** (Color online) Histogram of interatomic separations for (a) *P*-1 $MgLi_{10}$ at 100 GPa; (b) *Pm*-3*m* MgLi at 200 GPa; (c) *Pnma* MgLi at 500 GPa; (d) *I4/mmm* $Mg_3Li_2$ at 100 GPa; (e) *I4/mmm* $Mg_2Li$ at 300 GPa; (f) *P4/mmm* $Mg_3Li$ at 400 GPa; (g) *I4/mmm* $Mg_4Li$ at 400 GPa; (h) *P*-3*m*1 $MgLi_2$ at 500 GPa; (i) *P4/mmm* $Mg_5Li$ at 500 GPa, respectively. Note that those structures derived from BCC phase (b, d, e-g, i) have similar histogram.

The characteristics of newly predicted structures can be described by their interatomic separation histograms, as shown in Fig. S4. In $MgLi_{10}$, the coordination numbers of Li atoms vary in the ranges 6−10 with distances from 1.5Å to 2.5Å. Li atoms in MgLi-*Pm*-3*m* have coordination number 14, which is similar to bcc-Li. Li atoms in MgLi-*Pnma* are connected with six Mg atoms with a separation distance less than 1.75 Å as the nearest neighboring shell; and just above 2 Å, Li atoms coordinate with twelve other Li atoms and four Mg atoms. In $Mg_3Li_2$, Li atoms have coordination number 13, they are connected with eight Mg atoms in the nearest shell and five Li atoms in the second shell. The averaged local atomic environment of Li atoms in $Mg_2$Li-*I*4*/mmm* is similar to that of $Mg_3$Li-*P*4*/mmm* and $Mg_4$Li-*I*4*/mmm*, it contains a body-centered motif with coordination numbers of Li atoms equal to 8 (cubic coordination).

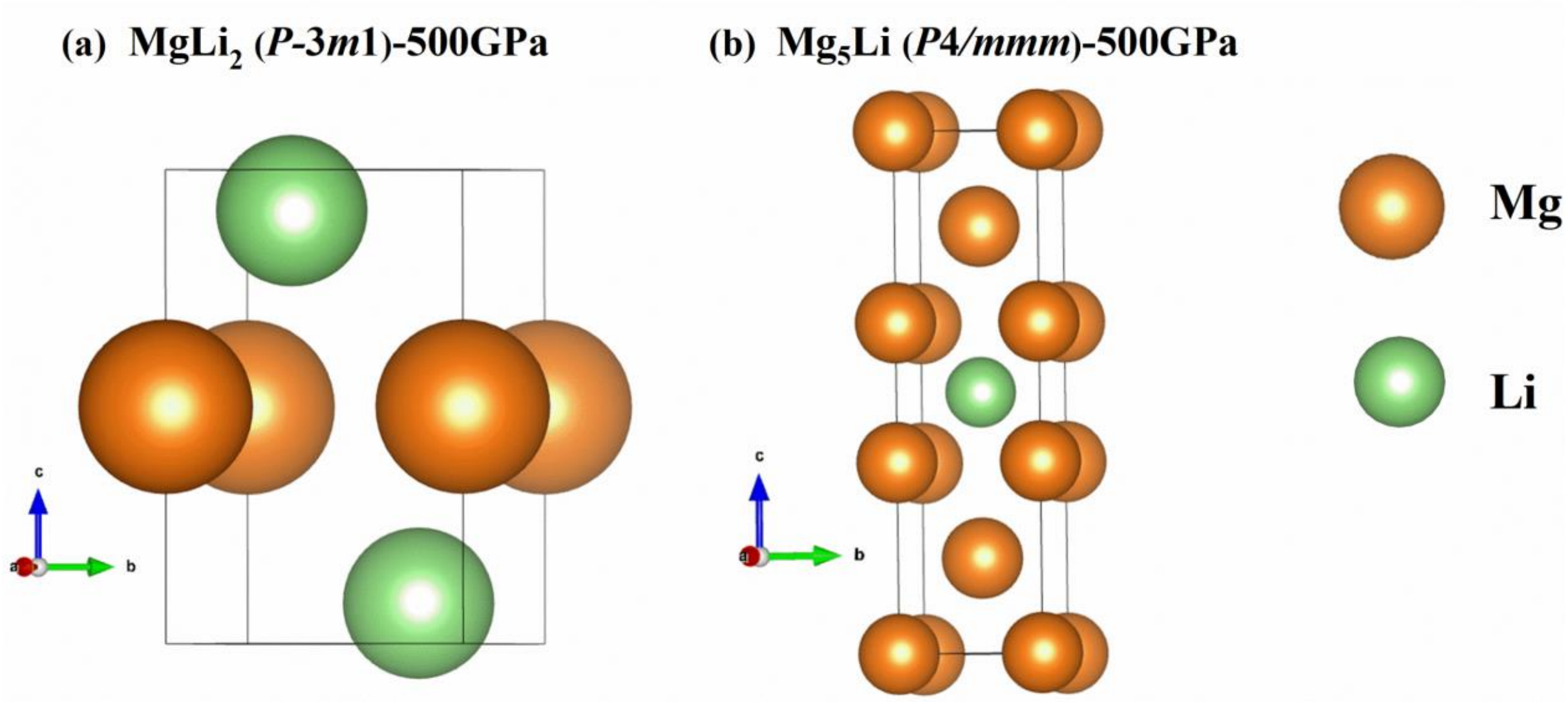

**Figure S5.** (Color online) (a) Structure of metastable *P*-3*m*1 $MgLi_2$ at 500 GPa; (b) Structure of metastable *P*4*/mmm* $Mg_5$Li at 500 GPa; Orange and green spheres represent Mg and Li atoms, respectively.

The new structure $MgLi_2$ and $Mg_5$Li are dynamically stable at 170-500 GPa and 0-500 GPa, respectively. As shown in Fig. S5, $MgLi_2$ is found to be $CdI_2$-type[2] structure (hP3 in the Pearson's symbol) and crystallizes in the trigonal *P*-3*m*1 space group. As shown in Fig. S4(h),

Li atoms are bonded to three Li atoms and three Mg atoms, the bond lengths for Li-Li and Li-Mg are 1.34 Å and 1.82Å, respectively. The Li atoms are located at body center and bonded with eight equivalent Mg atoms in *P*4/*mmm* $Mg_5Li$, and the Li-Mg cubic geometry inserted into the two-layer Mg cubic lattice as a guest. The coordination environment is similar to $Mg_2Li$-*I*4/*mmm*, and the Li–Li, Li–Mg1, and Li–Mg2 bonds are 2.14 Å, 1.83 Å, and 2.44 Å, respectively.

**Table S1.** Lattice parameters, atomic coordinates and Wyckoff site occupation of *P*-1 $MgLi_{10}$, *P*-3*m*1 $MgLi_2$, *Pm*-3*m* MgLi, *Pnma* MgLi, *I*4/*mmm* $Mg_3Li_2$, *I*4/*mmm* $Mg_2Li$, *P*4/*mmm* $Mg_3Li$, *I*4/*mmm* $Mg_4Li$, and *P*4/*mmm* $Mg_5Li$ at the given pressure.

| Phase | Lattice parameters (Å) | Atom | Site | Atomic coordinates |
|---|---|---|---|---|
| *P*-1 $MgLi_{10}$ (100 GPa) | *a*= 4.2731 *b*= 4.3904 *c*= 4.3942 *α*=87.97° *β*=88.97° *γ*=61.24° | Li | *1a* | (0.34099 0.88704 0.72062)<br>(0.05131 0.67931 0.88126)<br>(0.76970 0.00255 0.59576)<br>(0.11877 0.23071 0.46101)<br>(0.79363 0.60177 0.44728)<br>(0.39365 0.28971 0.17849)<br>(0.42905 0.37526 0.68948)<br>(0.39605 0.64581 0.32254)<br>(0.49193 0.90920 0.03875)<br>(0.93797 0.92672 0.23680) |
| | | Mg | *1a* | (0.85923 0.33964 0.95506) |
| *P*-3*m*1 $MgLi_2$ (500 GPa) | *a*=*b*=2.1069 *c*=3.2546 *α*=*β*= 90° *γ*=120° | Li | *2d* | (0.66667 0.33333 0.91395) |
| | | Mg | *1b* | (0.00000 0.00000 0.50000) |
| *Pm*-3*m* MgLi (200 GPa) | *a*=*b*=*c*=2.3955 *α*=*β*=*γ*=90° | Li | *1b* | (0.50000 0.50000 0.50000) |
| | | Mg | *1a* | (0.00000 0.00000 0.00000) |
| *Pnma* MgLi (500 GPa) | *a*= 3.3692 *b*= 3.3006 *c*= 3.3146 *α*=*β*=*γ*=90° | Li | *4c* | (0.30321 0.25000 0.51493) |
| | | Mg | *4c* | (-0.21068 0.25000 0.48872) |
| *I*4/*mmm* $Mg_3Li_2$ (100 GPa) | *a*=*b*=2.6098 *c*=13.5705 *α*=*β*=*γ*=90° | Li | *4e* | (0.50000 0.50000 0.09576) |
| | | Mg | *2a* | (0.00000 0.00000 0.00000) |
| | | | *4e* | (0.50000 0.50000 0.68865) |
| *I*4/*mmm* $Mg_2Li$ (300 GPa) | *a*=*b*=2.3127 *c*=7.3017 *α*=*β*=*γ*=90° | Li | *2a* | (0.00000 0.00000 0.00000) |
| | | Mg | *4e* | (0.00000 0.00000 0.35126) |
| *P*4/*mmm* $Mg_3Li$ (400 GPa) | *a*=*b*=2.2241 *c*=4.9553 *α*=*β*=*γ*=90° | Li | *1b* | (0.00000 0.00000 0.50000) |
| | | Mg | *1a* | (0.00000 0.00000 0.00000) |
| | | | *2h* | (0.50000 0.50000 0.71065) |
| *I*4/*mmm* $Mg_4Li$ (400 GPa) | *a*=*b*=2.2196 *c*=12.7959 *α*=*β*=*γ*=90° | Li | *2a* | (0.50000 0.50000 0.50000) |
| | | Mg | *4e* | (0.50000 0.50000 0.91775) |
| | | | *4e* | (0.50000 0.50000 0.30575) |
| *P*4/*mmm* $Mg_5Li$ (500 GPa) | *a*=*b*=2.1437 *c*=7.6985 *α*=*β*=*γ*=90° | Li | *1d* | (0.50000 0.50000 0.50000) |
| | | Mg | *1a* | (0.00000 0.00000 0.00000) |
| | | | *2h* | (0.50000 0.50000 0.81727) |
| | | | *2g* | (0.00000 0.00000 0.63353) |

**Table S2.** The valence state of Mg and Li atoms and the Bader charge of ISQs in $Mg_mLi_n$ compounds.

| **Phase** | **Mg (per atom)** | **Li (per atom)** | **ISQ (e/site)** | **ISQ(e/cell)** |
|---|---|---|---|---|
| *P*-1 $MgLi_{10}$ (100 GPa) | +0.65 | +0.97 | / | 7.51 |
| *Pm*-3*m* MgLi (200 GPa) | +1.16 | +0.6 | ISQ1: 0.44; ISQ2: 0.14 | 1.76 |
| *Pnma* MgLi (500 GPa) | +1.27 | +0.56 | 0.92 | 7.32 |
| *I*4/*mmm* $Mg_3Li_2$ (100 GPa) | +1.18 | +0.65 | ISQ1: 0.55; ISQ2: 0.64 ISQ3: 0.68; ISQ4: 0.50 | 9.71 |
| *I*4/*mmm* $Mg_2Li$ (300 GPa) | +1.20 | +0.56 | ISQ1: 0.64; ISQ2: 0.84 ISQ3: 0.64 | 5.92 |
| *P*4/*mmm* $Mg_3Li$ (400 GPa) | +1.20 | +0.53 | ISQ1: 1.11; ISQ2: 0.89 ISQ3: 0.61 | 4.11 |
| *I4/mmm* $Mg_4Li$ (400 GPa) | +1.20 | +0.53 | ISQ1: 1.15; ISQ2: 0.85<br>ISQ3: 0.61 | 10.64 |

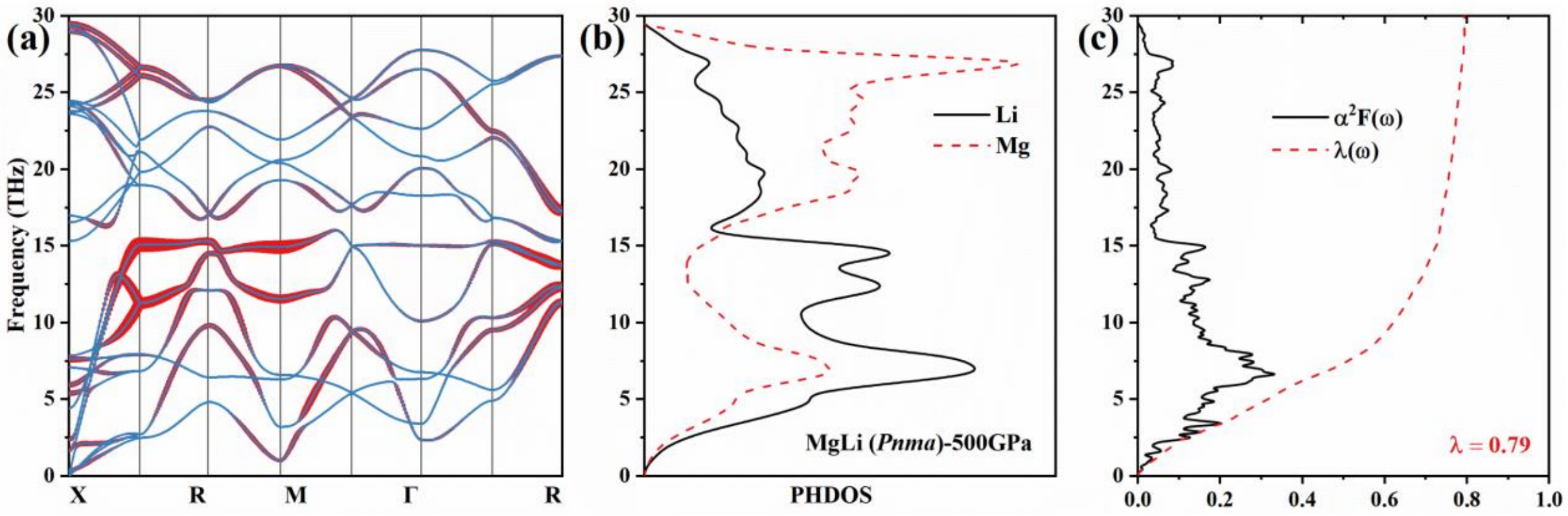

**Figure S6.** (Color online) (a) Phonon dispersion relations, (b) projected phonon density of states (PHDOS), (c) Eliashberg spectral function *α²F(ω)* and electron-phonon coupling constant *λ* of *Pnma* MgLi at 500 GPa. The sizes of the solid red circles are proportional to the electron phonon coupling strength.

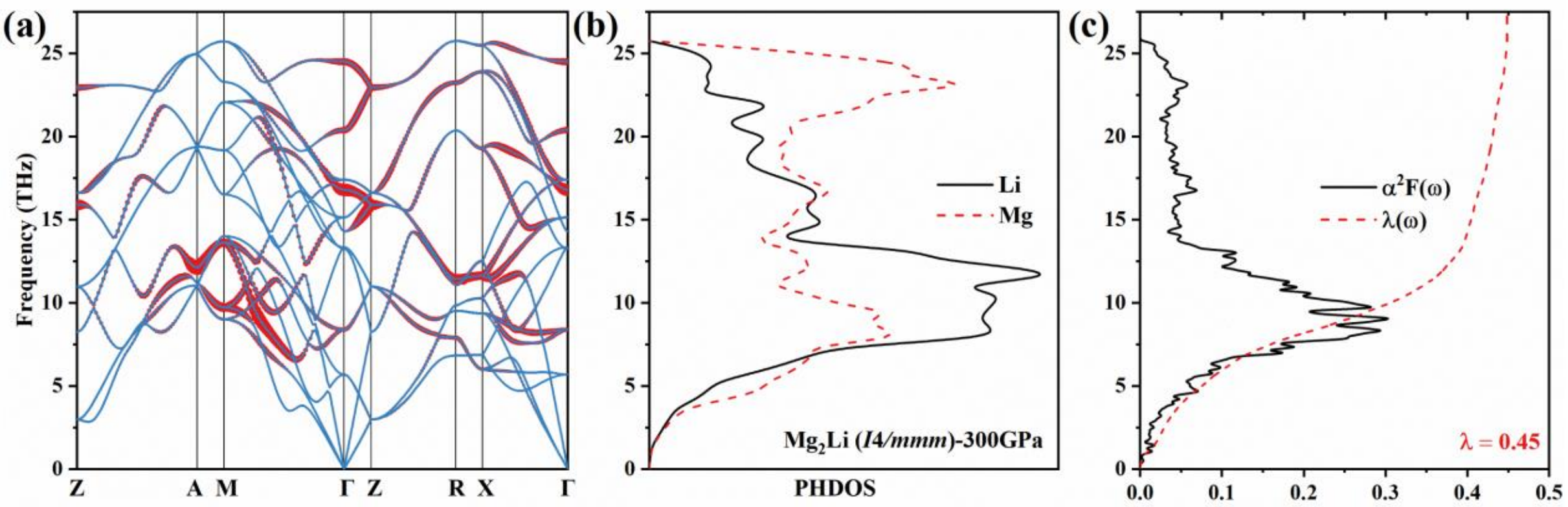

**Figure S7.** (Color online) (a) Phonon dispersion relations, (b) projected phonon density of states (PHDOS), (c) Eliashberg spectral function *α²F(ω)* and electron-phonon coupling constant *λ* of *I4/mmm* $Mg_2Li$ at 300 GPa. The sizes of the solid red circles are proportional to the electron-phonon coupling strength.

Figures S6 and S7 show the role of phonon dispersion on superconductivity. The calculated Eliashberg spectral function reflects a 77.7% contribution of phonons below 16 THz to the EPC constant, implying that the superconductivity of the electride MgLi-*Pnma* is dominated by low-frequency phonons. In MgLi-*Pnma*, the calculated EPC parameter (*λ*) is 0.79, and the logarithmic average frequency $\omega_{log}$ is 176.408, which corresponds to a $T_c$ of 8.91 K. Similarly, the low-frequency phonons below 15 THz play a dominant role in the superconductivity for $Mg_2Li$, due to their large contribution to $\lambda$ (78.8%). We conclude that low-frequency phonons jointly contributed by Li and Mg atoms, and the PHDOS of the Li atom is more concentrated at lower frequencies than Mg atoms.

**Table S3.** The charge state of ISQs and valence electron, *q/Q*, EPC parameter $\lambda$, and superconducting critical temperatures $T_c$ of typical electrides that contain Li.

| Phase | *q* (e/cell) | *Q* (e/cell) | *q/Q* | $\lambda$ | $T_c$ (K) |
|---|---|---|---|---|---|
| $Li_8Au$-*Fm*-3*m*-250 GPa | 3.0496 | 76 | 0.04 | 2 | 68.5[3] |
| $Li_5N$-*P6/mmm*-150 GPa | 1.2475 | 10 | 0.12475 | 1.39 | 48.97[4] |
| MgLi-*Pm*-3*m*-200 GPa | 1.76 | 3 | 0.58667 | 1.35 | 22.8 |
| $Li_{10}Se$-*C*2/*m*-50 GPa | 9.8228 | 32 | 0.30696 | 1.28 | 16[5] |
| $Li_8Au$-*Fm*-3*m*-200 GPa | 3.2044 | 76 | 0.04216 | 1.26 | 51.5[3] |
| $Li_5C$-*P6/mmm*-210 GPa | 0.5838 | 9 | 0.06487 | 1.26 | 48.3[6] |
| LiCa-*Fd*-3*m*-80 GPa | 3.8592 | 24 | 0.1608 | 1.25 | 18.9[7] |
| $Li_6As$-*C*2/*c*-270 GPa | 0.92 | 44 | 0.02091 | 1.2 | 45.2[8] |
| $Li_{10}Sb$-*C*2/*m*-100 GPa | 3.52 | 30 | 0.11733 | 0.8 | 9.2[8] |
| MgLi-*Pnma*-500 GPa | 7.32 | 12 | 0.61 | 0.79 | 8.91 |
| $Li_{11}Sb_2$-*C*2/*m*-300 GPa | 0.8772 | 42 | 0.02089 | 0.59 | 9.4[9] |
| $Li_8As$-*P6/mmm*-50 GPa | 3.27 | 13 | 0.25154 | 0.56 | 6.3[8] |
| Li-*Im*-3*m*-30 GPa | 1.4406 | 2 | 0.7203 | 0.51 | 14[10] |
| $Li_9Te$-*C*2/*m*-50 GPa | 7.6912 | 30 | 0.25637 | 0.5 | 4.01[11] |
| Li-*Im*-3*m*-20.3 GPa | 1.4878 | 2 | 0.7439 | 0.49 | 5.47[10] |
| Li10Te-*C*2/*m*-100 GPa | 3.51 | 32 | 0.10969 | 0.48 | 4[8] |
| Li-*Im*-3*m*-10 GPa | 1.5392 | 2 | 0.7696 | 0.47 | 2.3[10] |
| $Mg_2Li$-*I*4/*mmm*-300 GPa | 5.92 | 10 | 0.592 | 0.45 | 3.24 |

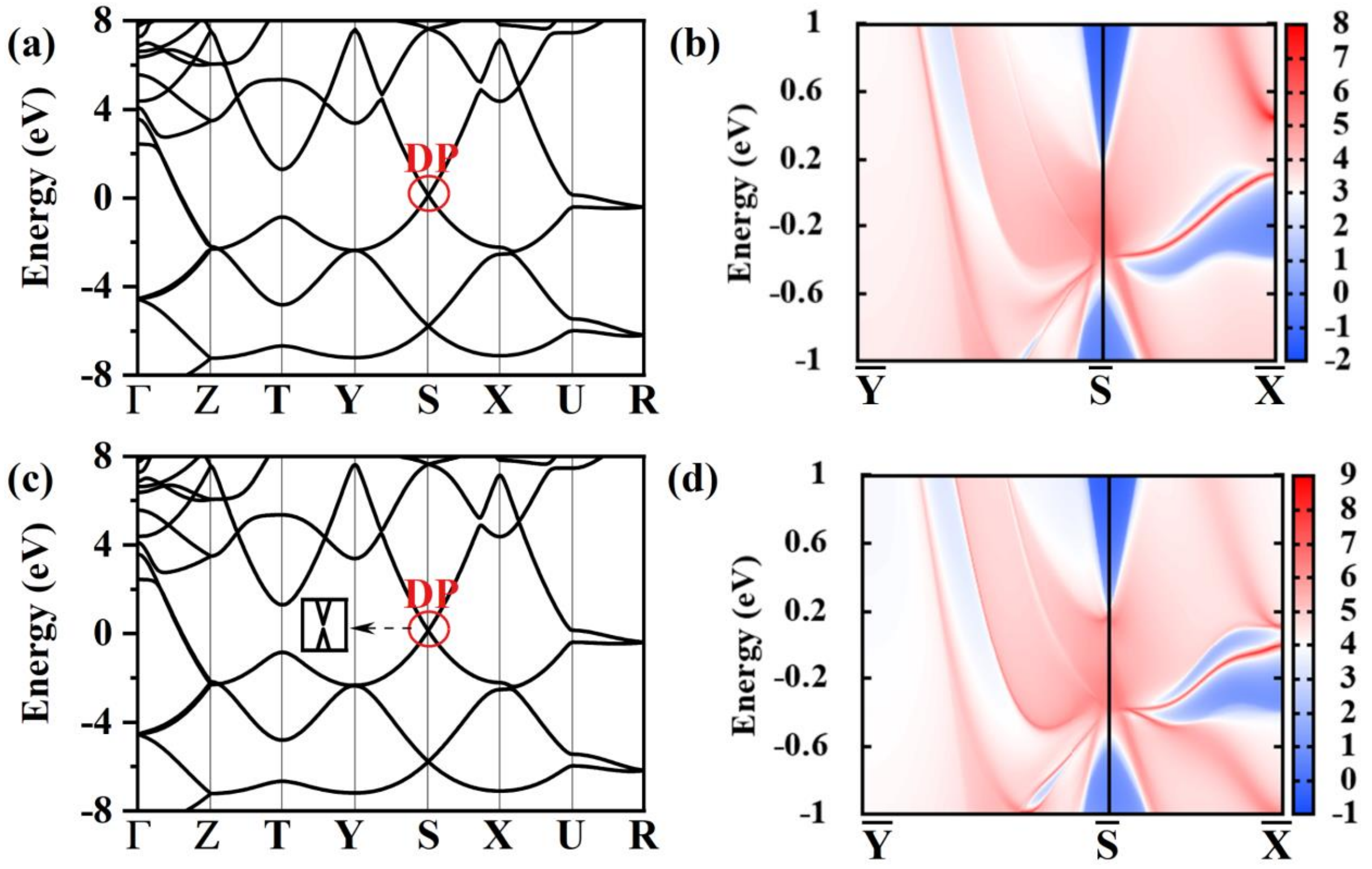

**Figure S8.** (Color online) (a) Band structure calculated without the SOC effect for MgLi-*Pnma*; (b) Surface states on the (001) surface of MgLi-*Pnma* without the SOC effect; (c) Band structure calculated with the SOC effect for MgLi-*Pnma*, the inset shows a zoomed-in view of the plot of the bands at DP with a gap; (d) Surface states on the (001) surface of MgLi-*Pnma* with the SOC effect; the Fermi level is set to zero. The pressure is 500 GPa.

**Table S4.** Parity products at TRIMs and the $Z_2$ Index of MgLi-*Pnma* at 500 GPa.

| TRIM | Parity Products | $v$ | $Z_2$ |
|---|---|---|---|
| Y | - | $v_1$=0, $v_1'$=1 | |
| 3S | + | $v_2$=0, $v_2'$=1 | |
| 3X | - | $v_3$=0, $v_3'$=1 | (1;111) |
| U | - | $v_0$=1 | |

We report a discovery of nontrivial $Z_2$ topology in the electronic structures of superconducting electride MgLi-*Pnma*, which is similar to MgLi-*Pm*-3*m* where the band-crossing point (Dirac point) is located at S points near the Fermi level and gapped with SOC effect. In Table S4 and Fig. S8, we find a topological $Z_2$ invariant of (1; 111) for the bulk band and topologically protected surface states in the (001) surfaces, signifying its nontrivial electronic topology.

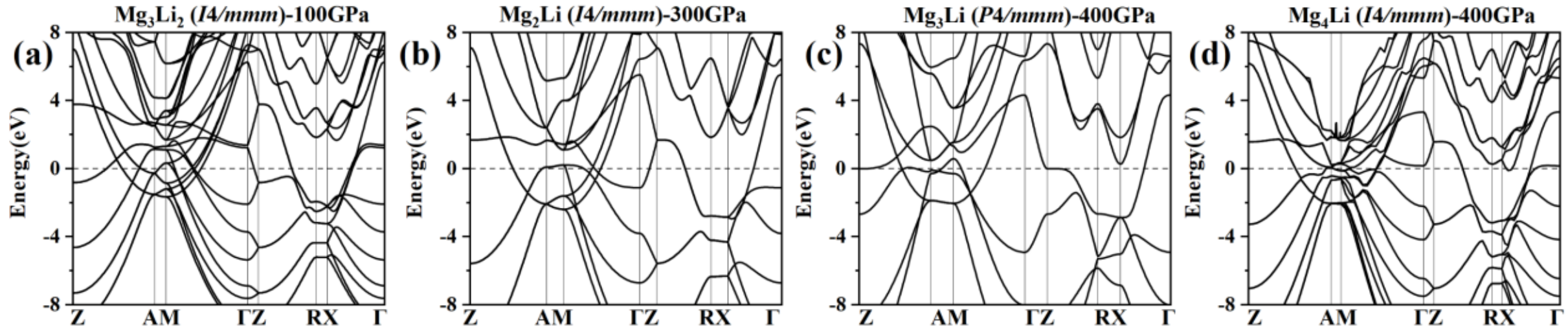

**Figure S9.** (Color online) Band structures for (a) *I*4/*mmm* $Mg_3Li_2$ at 100 GPa; (b) *I*4/*mmm* $Mg_2Li$ at 300 GPa; (c) *P*4/*mmm* $Mg_3Li$ at 400 GPa; (d) *I*4/*mmm* $Mg_4Li$ at 400 GPa, respectively.

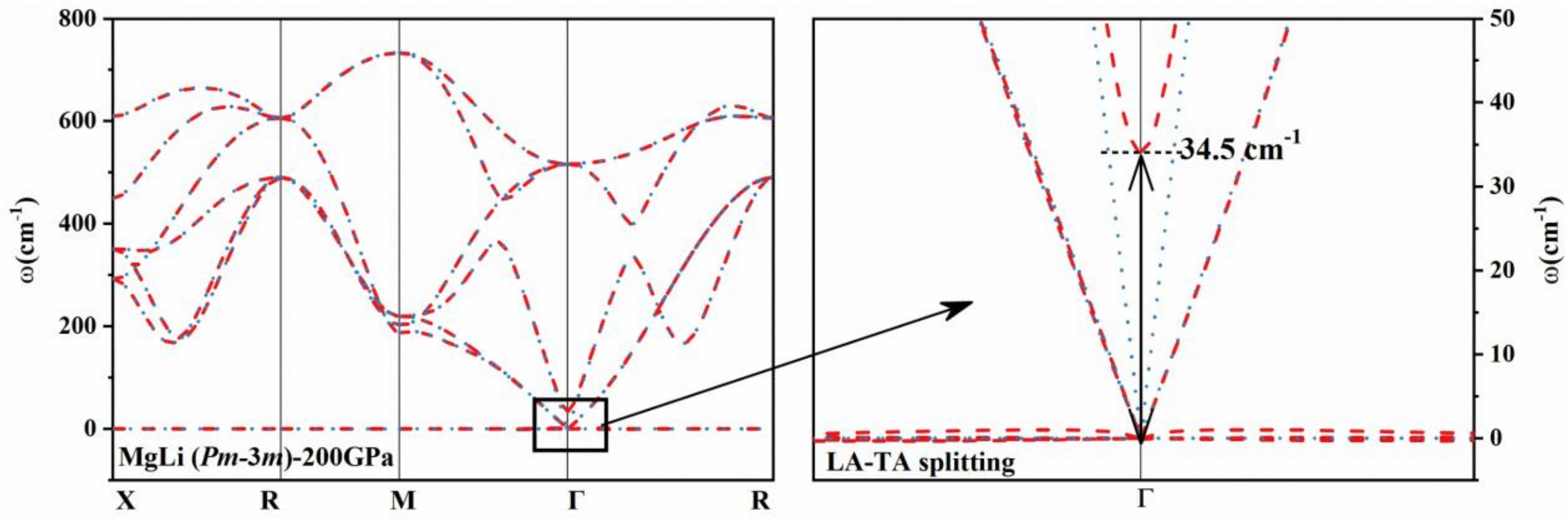

**Figure S10.** (Color online) Phonon dispersions of *Pm*-3*m* MgLi at 200 GPa, which are calculated using Bader charges (dot-dash-ed lines), and compared to that without taking nucleus-ISQ polarization into account (dotted lines in the right panel).

## Supplementary references